\documentclass[aps,prd,nofootinbib,preprintnumbers]{revtex4-2}
\usepackage{tikz}
   \usetikzlibrary{decorations.pathmorphing,positioning}
\usepackage{comment}
\usepackage{amsmath,amssymb}
\usepackage{graphicx}
\usepackage{hyperref}
\usepackage{bbold}
\usepackage[utf8]{inputenc}
\usepackage{slashed}
\usepackage{moresize}
\usepackage{braket}
\usepackage{geometry}

\newcommand{\bL}{\begin{Large}}
\newcommand{\eL}{\end{Large}}

\newcommand{\bea}{\begin{eqnarray}}
\newcommand{\eea}{\end{eqnarray}}
\newcommand{\be}{\begin{equation}}
\newcommand{\ee}{\end{equation}}

\usepackage[colorinlistoftodos]{todonotes}
\usepackage{appendix}
\newcommand{\cbl}[1]{{\textcolor{black}{#1}}}

\begin{document}

\preprint{\leftline{KCL-PH-TH/2026-{\bf 16}}}

\title{The $\omega$-Effect from a Multimode Squeezed Graviton State}

\author{Nick E. Mavromatos$^{a,b}$}
\author{Sarben Sarkar$^b$}
\medskip
\affiliation{$^a$Physics Division, School of Applied Mathematical and Physical Sciences,
National Technical University of Athens, Zografou Campus, Athens 157 80, Greece}
\affiliation{$^b$Theoretical Particle Physics and Cosmology Group, Department of Physics,
King's College London, London, WC2R 2LS, UK}

\date{\today}

\begin{abstract}
The $\omega$-effect in entangled neutral-meson systems provides a sensitive
probe of CPT violation induced by quantum-gravitational environments. In
open quantum systems, interactions with inaccessible gravitational degrees
of freedom can render the reduced meson dynamics non-unitary, causing the
CPT operator to become ill-defined, even when the underlying microscopic
Hamiltonian is CPT invariant. We present a microscopic derivation of the
$\omega$-effect arising from a multimode squeezed gravitational environment
generated by an axion cloud around a Kerr black hole. Using the Takagi
decomposition of the associated complex symmetric squeezing kernel, the graviton field
is expressed in terms of independent squeezed supermodes
$\mathbb{b}_\alpha$ possessing anomalous correlators
$\langle \mathbb{b}_\alpha \mathbb{b}_\alpha \rangle \neq 0$. These
correlators provide a microscopic quantum counterpart of the stochastic
fluctuations that appear in earlier D-particle foam descriptions of the
$\omega$-effect, replacing phenomenological variances of D-particle recoil by calculable
graviton correlation functions. The anomalous gravitational correlators induce correlated
flavour transitions in a neutral-kaon system whose propagation eigenstates
$K_S$ and $K_L$ arise from 
$K^0$--$\bar K^0$ mixing induced by the weak interactions. After tracing over the
graviton bath, the anomalous correlators and the mixing combine to generate transitions between the
antisymmetric and symmetric two-meson sectors. This results in a small exchange-symmetric
admixture, parametrised by $\omega$, in the otherwise antisymmetric EPR
state. We obtain an explicit expression for $\omega$ in terms of a sum over Takagi
supermodes weighted by their squeezing amplitudes and phases together with
the weak-interaction flavour-mixing matrix element. The resulting framework suggests that the $\omega$-effect may be a generic
signature of non-classical states of gravitational environments, extending beyond
the specific axion-cloud scenario considered here. The observability of
the $\omega$-effect from other astrophysical and microscopic black-hole sources
is discussed.

\end{abstract}

\maketitle

\section{Introduction}
\label{sec:intro}

The question of whether CPT symmetry is an exact and universal microscopic symmetry of nature stands as one of the major open problems at the interface of quantum mechanics and quantum field theory \cite{Mavromatos:2005bg,Kostelecky:1991ak}. In quantum field theory CPT invariance is
guaranteed for theories possessing Lorentz invariance,
locality, and unitarity~\cite{Luders:1954zz,Pauli:1955}. When 
gravitational effects are taken into account, these assumptions are not valid. Assuming gravity can be quantised, (topologically non-trivial) spacetime fluctuations at the Planck scale (spacetime foam~\cite{Wheeler:1998vs}) may require the description of
matter in terms of \emph{open quantum systems} in a reduced state space. Such systems interact with a gravitational
environment, causing unitarity of the reduced dynamics to be lost. Because the
CPT operator is antiunitary---it contains the time-reversal operator $T$, which
must be antiunitary to preserve the Schr\"odinger equation under the transformation $t\to
-t$~\cite{Wigner1959}---non-unitary evolution prevents the consistent
definition of CPT on the reduced Hilbert space~\cite{Wald:1980,Ellis:1983jz}.

A sensitive probe for this effect is provided by entangled neutral-meson
systems produced at meson factories~\cite{KLOE:2006}. A
$\phi$-factory, such as
DA$\Phi$NE at Frascati~\cite{Amelino-Camelia:2010cem}, is an electron--positron collider operating at the $\phi(1020)$-meson
resonance ( $m_\phi\simeq 1020\,\text{MeV}$). At such a facility the decay $\phi\to K^0\bar{K}^0$
produces a $K^0\bar{K}^0$ pair in a $P$-wave ($L=1$); since the $\phi$ has
$J^{PC}=1^{--}$ and the kaons are spin-zero pseudoscalars, Bose symmetry requires
the flavour state to be exchange-antisymmetric:
\begin{equation}
|A\rangle
= \frac{1}{\sqrt{2}}
\bigl(
|K^0(\mathbf{k})\rangle|\bar{K}^0(-\mathbf{k})\rangle
-
|\bar{K}^0(\mathbf{k})\rangle|K^0(-\mathbf{k})\rangle
\bigr).
\end{equation}
If CPT is well-defined, the corresponding exchange-symmetric state $|S\rangle$
is strictly forbidden. As argued by Bernab{e}u, Mavromatos, and
Papavassiliou~\cite{Bernabeu:2004prl} and demonstrated explicitly in a toy
model of stochastic quantum gravity in~\cite{Bernabeu:2006}, the ill-defined nature of the CPT operator~\cite{Wald:1980}, in open
quantum gravitational systems, \cbl{is due to the decoherence of propagating ordinary  matter at low-energy (compared to the Planck scale) \cite{Mavromatos:2004sz,Sarkar:2009jw} }. 
This allows the initial state of the entangled neutral mesons
to acquire a small symmetric admixture denoted by $|S\rangle$,
\begin{equation}
  |\psi\rangle = |A\rangle + \omega\,|S\rangle\,.
  \label{eq:omega_def}
\end{equation}
\cbl{In Eq.~\eqref{eq:omega_def}}, the \emph{complex} parameter $\omega$ is a direct observable signature of CPT
violation, measuring the $\omega$-\emph{effect}~\cite{Bernabeu:2004prl}.
Experimental bounds $|\omega| \lesssim 10^{-3}$ have been established at \cbl{neutral-Kaon
(KLOE)~\cite{KLOE:2006,KLOE-2:2021}) and other neutral-meson}
facilities~\cite{Alvarez:2004tj,Alvarez:2006ry,Bernabeu:2016kva,Shi:2016bvo},
with independent, weaker bounds from BESIII~\cite{Prasad:2023uwe}.

The existing phenomenological model for $\omega$ is based on a D-foam
\cite{Bernabeu:2006,Ellis:1997jw,Ellis:2000sx,Ellis:2004ay,Mavromatos:2009pp,Sarkar:2009jw},
in which spacetime zero-dimensional brane defects (D particles~\cite{pol1,pol2,Szabo20111}) scatter kaons and flip their internal quantum
numbers by deforming the gravitational metric. As a neutral meson propagates through this fluctuating medium, each
scattering induces a tiny fluctuation in the effective metric \cite{Bernabeu:2006,Sarkar:2009jw}, that can  induce flavour transitions. The model does not specify the quantum state of the gravitational
environment These effects are encoded in a
phenomenological Hamiltonian in terms of classical stochastic random variables $r_i$ with vanishing mean and
non-vanishing variance $\langle r_i r_j\rangle = \Delta_i\delta_{ij}$, which
measure the strength of the spacetime fluctuations~\cite{Bernabeu:2006,Mavromatos:2009pp,Sarkar:2009jw}.
The stochastic Hamiltonian
\begin{equation}
\widehat H_I = -(r_1\sigma_1+r_2\sigma_2)\,\hat k,
\label{eq:HI_stochastic}
\end{equation}
with off-diagonal Pauli matrices $\sigma_{1,2}$ acting in a two-dimensional flavour space,
induces transitions between the two neutral-meson states.
In flavour space, applying
Rayleigh-Schr\"{o}dinger perturbation theory, off-diagonal Pauli-matrix
couplings  dress the meson eigenstates and add  to the
  antisymmetric EPR state \cite{Nielsen:2012yss},  a symmetric state $S$ with 
\begin{align}\label{omestim}
|\omega|^2\sim \frac{\Delta_2 k^2}{(m_1-m_2)^2}\,,\end{align}
\cbl where $k$ is the magnitude of the $3$ momentum in a centre of mass frame of the decaying $\phi$ particle and $m_1$ and $m_2$ are the masses of the propagating eigenstates.
\cbl{The magnifying factor $(m_1-m_2)^{-1}$, due to the tiny mass difference between the neutral-meson energy eigenstates, counteracts the quadratic Planck mass suppression induced by the stochastic variance  $\Delta_2$;  the latter represents 
effects due to quantum-gravity environments~\cite{Bernabeu:2006}.}
In the D-foam recoil estimate, the stochastic variance is parametrised as
\begin{equation}
\Delta_2
=
\langle r_2^2\rangle
\sim
\zeta^2\frac{k^2}{M_{\rm QG}^2}.
\end{equation}
If the quantum-gravity scale is identified with the Planck mass,
\(M_{\rm QG}\sim M_{\rm Pl}\), this becomes
\begin{equation}
\Delta_2
\sim
\zeta^2\frac{k^2}{M_{\rm Pl}^2}.
\end{equation}
Thus \(\Delta_2\) is quadratically Planck-suppressed. Substitution into \eqref{omestim}, and taking the square root, 
yields
\begin{equation}\label{eq:omega_estimate}
|\omega|
\sim
\zeta
\frac{k^2}
{M_{\rm Pl}|m_1-m_2|}.
\end{equation} 
\cbl{By contrast to this model, we will derive here the environmental fluctuations from a specific
multimode squeezed graviton state generated by an axion cloud around a Kerr
black hole~\cite{Dorlis:2025zzz,Dorlis:2025amf}. This allows the $\omega$-effect
to be related directly to graviton correlation functions and Takagi supermodes \cite{Houde:2024mkj} associated with squeezed states.}

\cbl{
The decay of parent particles, \emph{e.g.} 
$\phi$-mesons in the entangled neutral-kaon case, 
inside the axionic cloud, supported by rotating (Kerr-type) black holes \cite{Dorlis:2025amf}, is a rare phenomenon; nonetheless it may pave the way for a microscopic understanding of possible $\omega$-effects in microscopic quantum-gravity environments \cite{Mavromatos:2009pp,Clifton:2011jh,Sarkar:2009jw} containing squeezed quantum-graviton fluctuations. We show that the latter can induce the $\omega$ effect for $\phi$-mesons, and parent particles of other types of entangled neutral mesons (such as $B$ and $D$ mesons).
\(\phi\)-mesons can be produced in a variety of high-energy astrophysical
and cosmological environments.
One possibility is through collisions of
ultra-high-energy cosmic rays with nuclei in the Earth's upper atmosphere
or with interstellar gas. Once the centre-of-mass energy exceeds the
production threshold, approximately \(2m_p+m_\phi\simeq 2.84~{\rm GeV}\)
for proton--proton collisions, \(\phi\)-mesons may be produced through
strong-interaction processes such as
\[
pp\rightarrow pp\phi,
\qquad
pN\rightarrow pN\phi.
\]
A second source is provided by core-collapse supernovae, where
temperatures of order tens of MeV and baryon densities near or above
nuclear density create a hadronic environment in which strange mesons,
including \(\phi\)-mesons, can be copiously produced through
nucleon-nucleon and meson-meson interactions. Finally, in the early
Universe, during the hadronisation epoch following the quark-gluon
plasma phase (\(T\sim100\!-\!200~{\rm MeV}\)), strange quarks (s) and
antiquarks are abundant, leading to efficient production of strange
hadrons, including \(\phi(s\bar s)\) mesons, before chemical freeze-out.
In such environments, squeezed graviton states~\cite{Dorlis:2026gth} can contribute to the appearance of $\omega$-effect-like situations through  $\phi$-meson 
decays to entangled neutral kaons \emph{etc.}.} 

The rotating black-holes that we consider can be  astrophysical or primordial in origin.  
\cbl{The results of the present work may also have implications for
analogue-gravity systems \cite{Jacquet:2020bar,Barcelo:2005fc}. The essential ingredient
in our mechanism is not the astrophysical black hole itself, but the
generation of a non-classical squeezed environment through pair-production
processes in a rotating background. Similar Bogoliubov transformations
arise in a variety of analogue-gravity platforms, including fluid,
optical, and condensed-matter realisations of black-hole horizons, where
correlated pairs of quasiparticles are produced and the outgoing state is
naturally squeezed. In this sense, analogue systems may provide
laboratory realisations of some of the intermediate steps of the
black-hole--axion-cloud mechanism. While such systems do not reproduce
the neutral-kaon sector or CPT violation itself, they may offer an
experimental arena in which the generation of squeezed environmental
states and their influence on coupled quantum subsystems can be studied.
This suggests a possible connection with ongoing analogue-gravity
programmes investigating Hawking radiation, superradiance, and
quantum-correlation effects in horizon analogues. While an ultimate objective is to analyze the possibility of squeezed quantum-graviton states within microscopic spacetime-foam environments~\cite{Wheeler:1998vs}, such an investigation lies beyond the scope of this study.}

\color{black}The paper is organised as follows. Section~\ref{sec:squeezed} constructs the
multimode squeezed graviton state and performs the Takagi decomposition. It also
establishes in Section~\ref{sec:stochastic_dict} a  correspondence between
the stochastic D-foam variance and the anomalous graviton correlators of the
present approach. Section~\ref{sec:interaction} derives the reduced meson
dynamics to second order in the meson--graviton coupling.
Section~\ref{sec:omega} contains the central quantitative result: the
expression for the $\omega$-parameter as a sum over Takagi supermodes weighted
by the anomalous squeezed-state correlator, together with its near-resonant and
Markovian limits and the open-system derivation of CPT violation.
An application to the neutral kaon system is given in Section~\ref{sec:kaon},
and an order-of-magnitude estimate, including a discussion of astrophysical
source dilution, appears in a subsection~\ref{sec:estimate}. We conclude in
Section~\ref{sec:conclusions}. Technical aspects of our approach, and background material, are presented in several Appendices.

\section{Multimode Squeezed Graviton State}
\label{sec:squeezed}

\subsection{Linearised gravity and the axion--graviton coupling}

A gravitational environment \cite{Dorlis:2025amf,Dorlis:2026gth,Dorlis:2025zzz} (coupled to neutral mesons) need not be specified only by its classical
metric perturbation. At the quantum level the graviton may be in
different states such as the vacuum, a thermal state, a coherent state
corresponding to a classical gravitational wave, and a squeezed state
containing correlated graviton pairs\footnote{Beyond vacuum, thermal, coherent and squeezed states, the gravitational
environment may in principle occupy more general non-Gaussian states.
Such states possess higher-order connected correlation functions that are
not determined solely by two-point correlators and therefore cannot be
fully characterised by a covariance matrix
\cite{Genoni:2008}.}. 
These possibilities have different implications for the reduced
meson dynamics (obtained after tracing over the environmental degrees of freedom). Vacuum, thermal and coherent states possess only normal
two-point functions and therefore generate ordinary phase shifts,
decoherence, absorption and emission terms. By contrast, a squeezed
graviton state has anomalous pair correlators for gravitational destruction operators $ \mathbb b_\alpha$ (where $\alpha$ labels states of the graviton)
\begin{align}\label{bcorr}
\langle \mathbb b_\alpha \mathbb b_\beta\rangle \neq 0 ,
\end{align}
which are precisely the correlations required for the simultaneous
two-meson transitions that generate the \(\omega\)-effect. \cbl{In \eqref{bcorr} the expectation value $\langle \dots \rangle$ is taken with respect to the appropriate squeezed vacuum states, which are \emph{not} annihilated by the $\mathbb b_\alpha$ operators~\cite{ScullyZubairy} (\emph{c.f.} Appendix \ref{app:takagi})}.\footnote{\cbl{The (multimode) squeezed vacuum $|\Psi\rangle$ is related to the ordinary Fock vacuum state $|0\rangle $, annihilated by destruction operators $a_i$, 
by means of an appropriate Bogolubov transformation, schematically $|\Psi\rangle = S_\Psi \, |0\rangle$, 
$S_\Psi  \propto \exp\Big(\frac{1}{2} \sum_{i,j} \mathcal G_{ij} \, a^\dagger_i \,  a^\dagger_j  - {\rm h.c.}\Big)$, where $a_i$ ($a_i^\dagger)$ are generic annihilation (creation) operators, such that $a_i 
|0\rangle =0$, and 
$\mathcal G_{ij}$ is the squeezing kernel (\emph{cf.} \eqref{eq:squeezed_vac}). The significant enhancement of the effects of a squeezed state can be readily seen already in the case of single-mode squeezing. Indeed, in that case the 
squeezing kernel reduces to a single complex parameter $\zeta=r \, \exp(i\theta)$, and the squeezing operators assume the form
$S(\zeta) = \exp\Big(\frac{1}{2} (\zeta^\star \, (a)^2 - \zeta\, (a^\dagger)^2)\Big)$, and are such that $S^\dagger(\zeta) = S^{-1}(\zeta)$ and $S^\dagger(\zeta) \, a \, S(\zeta) = a\, \cosh (r) - a^\dagger \, e^{-i\theta}\, \sinh (r)$, clearly leading to significant enhancement of the effects in the appropriate correlators for $r > 1$.}}

\vskip .3cm

An astrophysical example of such a state, is the gravitational radiation produced by an axion
cloud around a rotating (Kerr) black hole. This case was argued in \cite{Dorlis:2025zzz,Dorlis:2025amf,Dorlis:2026gth} 
to produce significant squeezing of polarisation-entangled quantum graviton states. This has ramifications for potential detection of such quantum-gravity signatures in gravitational waves at future interferometers, thereby opening a window for an experimental verification of quantum gravity (at least as an effective field theory). In this article we assume the production of $\phi$-mesons as a concrete way of producing entangled particle (neutral-kaon)  states, whose EPR correlators are afflicted by an induced $\omega$-effect~\cite{Bernabeu:2004prl}, as a result of their interaction with the squeezed quantum graviton states produced by the axions in the cloud. Because this environment involves
a quantum field, its influence on the meson subsystem is governed by
graviton two-point functions, and so we must quantise the gravitational
perturbations generated by the cloud before computing those correlators.

It is useful to contrast the squeezed graviton environment with a 
ordinary thermal gravitational bath. For a thermal state diagonal in the
number basis one has \cite{ScullyZubairy}
\begin{equation}
\langle \mathbb b_\alpha^\dagger \mathbb b_\beta\rangle
=
\delta_{\alpha\beta} n_\alpha,
\qquad
\langle \mathbb b_\alpha \mathbb b_\beta^\dagger\rangle
=
\delta_{\alpha\beta}(n_\alpha+1),
\end{equation}
but
\begin{equation}
\langle \mathbb b_\alpha \mathbb b_\beta\rangle
=
\langle \mathbb b_\alpha^\dagger \mathbb b_\beta^\dagger\rangle
=
0.
\end{equation}
Thus a thermal bath supplies only normal correlations.

Before delving into the detail of a neutral-kaon Hamiltonian, it is useful to
display the mechanism for the $\omega$-effect in a minimal effective theory. We treat each neutral $K$ meson
as a \emph{two-level} system spanned by the propagation states
\[
|K_S\rangle,\qquad |K_L\rangle .
\]
Weak interactions determine the propagation eigenstates
\(K_S\) and \(K_L\) through the mixing of \(K^0\) and
\(\bar K^0\). Once these single-particle states have been identified,
the relevant Hilbert space for the \(\omega\)-effect is the
two-kaon sector.

The states
\begin{equation}
|K_SK_L\rangle,
\qquad
|K_LK_S\rangle,
\end{equation}
span a two-dimensional subspace that may be decomposed into
eigenstates of the exchange operator \(P_{12}\),
which interchanges the two mesons.
The antisymmetric and symmetric combinations are
\begin{equation}
|A\rangle
=
\frac{1}{\sqrt2}
\left(
|K_SK_L\rangle
-
|K_LK_S\rangle
\right),
\qquad
P_{12}|A\rangle=-|A\rangle,
\end{equation}
and
\begin{equation}
|S\rangle
=
\frac{1}{\sqrt2}
\left(
|K_SK_L\rangle
+
|K_LK_S\rangle
\right),
\qquad
P_{12}|S\rangle=+|S\rangle.
\end{equation}

The state produced in
\(\phi\to K^0\bar K^0\)
is antisymmetric. The \(\omega\)-effect corresponds to the appearance
of an exchange-symmetric admixture,
\begin{equation}
|A\rangle
\longrightarrow
|A\rangle+\omega |S\rangle .
\end{equation}

The role of the weak interaction is therefore to define the
single-particle basis \((K_S,K_L)\), whereas the \(\omega\)-effect
probes mixing between the exchange-symmetry sectors of the resulting
two-particle state.

\color{black} The interaction with the \(\alpha\)-th graviton supermode is, for the moment, written
generically as
\begin{equation}
H_{\rm int}
=
\sum_{a=1}^{2}\sum_\alpha
\left[
g_\alpha^{(a)}X^{(a)}\mathbb b_\alpha
+
g_\alpha^{(a)*}X^{(a)}\mathbb b_\alpha^\dagger
\right],
\label{eq:Hint_schematic}
\end{equation}
where \(a=1,2\) labels the two mesons. The operator \(X^{(a)}\) acts only
on the \(a\)-th meson. Its microscopic origin arises from the projection of the
gravitational stress-tensor coupling onto the neutral-kaon sector, \emph{but
that derivation will be given in detail later}.

The only property of \(X\) needed at this stage is that, in the
\(\{|K_S\rangle,|K_L\rangle\}\) subspace, it  contains transition
terms
\begin{equation}
X_{\rm tr}
=
c_+\sigma_+
+
c_-\sigma_-,
\qquad
\sigma_+=|K_L\rangle\langle K_S|,
\qquad
\sigma_-=|K_S\rangle\langle K_L|.
\end{equation}
These terms raise or lower the meson between the two propagation
eigenstates.
The initial two-kaon state produced at a \(\phi\)-factory is
antisymmetric~\cite{KLOE:2006},
\begin{equation}
|A\rangle
=
\frac{1}{\sqrt2}
\left(
|K_SK_L\rangle
-
|K_LK_S\rangle
\right).
\end{equation}
The \(\omega\)-effect corresponds to the appearance of a symmetric
component~\cite{Bernabeu:2004prl}
\begin{equation}
|S_\omega\rangle
=
\frac{1}{\sqrt2}
\left(
|K_SK_L\rangle
+
|K_LK_S\rangle
\right).
\end{equation}
Equivalently, in terms of a meson density matrix $\rho_M$, we have a non-zero matrix element
\begin{equation}
\omega(t)
\sim
\langle S_\omega|\rho_M(t)|A\rangle .
\end{equation}

The terms relevant for mixing \(|A\rangle\) and \(|S\rangle\) are the
cross terms
\begin{equation}
\sigma_+^{(1)}\sigma_-^{(2)},
\qquad
\sigma_-^{(1)}\sigma_+^{(2)}.
\end{equation}
They act inside the
\(\{|K_S K_L\rangle,|K_L K_S\rangle\}\) subspace and therefore can
distinguish the antisymmetric and symmetric combinations.

We derive later that in the open-system calculation (where there is a trace over the gravitational degrees of freedom) these two-meson operators arise from the
second-order expression
\begin{equation}
H_{\rm int}(t_1)H_{\rm int}(t_2)
\supset
X^{(1)}(t_1)X^{(2)}(t_2)\,
\mathbb b(t_1)\mathbb b(t_2).
\end{equation}
After tracing over the graviton bath this gives
\begin{equation}
\langle \mathbb b(t_1)\mathbb b(t_2)\rangle\,
X^{(1)}(t_1)X^{(2)}(t_2).
\end{equation}

Thus the essential chain is
\begin{equation}
\boxed{
\langle \mathbb b\mathbb b\rangle
\neq 0
\quad\Longrightarrow\quad
X^{(1)}X^{(2)}
\quad\Longrightarrow\quad
|A\rangle \leftrightarrow |S\rangle
\quad\Longrightarrow\quad
\omega\neq0 .
}
\end{equation}

For a thermal, vacuum, or ordinary coherent graviton state one has
\begin{equation}
\left\langle
\mathbb b_\alpha(t_1)\mathbb b_\beta(t_2)
\right\rangle
=
0.
\end{equation}
Such states may generate ordinary decoherence, absorption, emission, or
phase shifts through normal correlators such as
\(\langle\mathbb b_\alpha^\dagger(t_1)\mathbb b_\beta(t_2)\rangle\), but
they do not generate the simultaneous two-meson transition channel
required for the \(\omega\)-effect in this mechanism.
\cbl{This is also in agreement with the considerations in \cite{Bernabeu:2006}, in which it is demonstrated that, generically, thermal environments do not induce an $\omega$-effect.}

By contrast, a squeezed graviton state has a non-vanishing anomalous
pair correlator,
\begin{equation}
\left\langle
\mathbb b_\alpha(t_1)\mathbb b_\beta(t_2)
\right\rangle
\neq 0.
\end{equation}
This is the basic reason why the gravitational environment considered
here must be a non-classical, pair-correlated graviton state rather than
an ordinary thermal or coherent gravitational background. The detailed
derivation of \(X\), of the interaction-picture phases, and of the full
open-system expression for \(\omega(t)\) is given in the following
sections.
Consequently a thermal graviton bath does not generate the
\(\omega\)-effect through the mechanism studied here (\cbl{see also discussion in \cite{Bernabeu:2006}}). This is why the
non-classical squeezed nature of the gravitational environment is
essential.

\vskip .3cm 

\cbl{In what follows, we proceed  to construct, in detail, the 
squeezed-quantum-gravitational $\omega$-effect.}
\cbl{We commence our discussion with the \emph{quantisation of gravity in the weak effective field theory limit}}~\cite{tHooft:1974toh,donogh,Rovelli_2004,Percacci:2017fkn} by
expanding the metric $g_{\mu\nu}$ as a small perturbation around the flat spacetime metric $\eta_{\mu\nu}$,
\footnote{We use standard gauge fixing with Fadeev--Popov ghosts;
graviton--graviton scattering is ignored.}
\begin{equation}
g_{\mu\nu} = \eta_{\mu\nu} + \kappa \, h_{\mu\nu},
\qquad \kappa = \sqrt{8\pi G} = 1/M_{\rm Pl}.
\end{equation}

\cbl{In our approach, the squeezed graviton states,  which induce the \(\omega \)-effect, arise in the astrophysical system of a rotating black hole with a superradiant axion-like cloud surrounding its exterior horizon; we consider Chern-Simons gravity~\cite{jackiw} as our effective field theory \cite{Dorlis:2025zzz,Dorlis:2025amf,Dorlis:2026gth}.
The pertinent effective action} for a pseudoscalar axion field  $b(x)$ of mass $\mu_b$ coupled
to the gravitational field is~\footnote{Our conventions and definitions used throughout this work are: signature of metric $(-, +,+,+ )$, Riemann Curvature tensor:
$R^\lambda_{\,\,\,\,\mu \nu \sigma} = \partial_\nu \, \Gamma^\lambda_{\,\,\mu\sigma} + \Gamma^\rho_{\,\, \mu\sigma} \, \Gamma^\lambda_{\,\, \rho\nu} - (\nu \leftrightarrow \sigma)$, Ricci tensor $R_{\mu\nu} = R^\lambda_{\,\,\,\,\mu \lambda \nu}$, and Ricci scalar $R_{\mu\nu}g^{\mu\nu}$. We also work in units $\hbar=c=1$.}`
\begin{equation}
S=\int d^4x \,\sqrt{-g}\,\left[\frac{R}{2\kappa^2}
-\frac{1}{2}(\partial_\mu b)(\partial^\mu b)
-\frac{1}{2}\mu^{2}_b b^2 - A\, b\,R_{CS} \right],
\label{eq:Action}
\end{equation}
where $R_{CS} = \frac{1}{2}R^{\mu}{}_{\nu\rho\sigma}
\widetilde{R}^{\nu\,\rho\sigma}{}_{\mu}$ is the gravitational Chern-Simons (CS)
anomaly\footnote{This term appears through the Green--Schwarz mechanism in
10-dimensional superstring theory~\cite{GS}, \cbl{from our action emanates in the low-energy limit relative to the string scale.}}, $R$ is the Ricci scalar and the string scale $A \sim 10^{-2} M_{\rm
Pl}/M_s$ in string theory~\cite{Duncan:1992vz}. In the absence of the CS term, the axion
stress-energy tensor is
\begin{equation}
T_{\mu\nu}^{(b)}
=
\nabla_\mu b\,\nabla_\nu b
-\frac12 g_{\mu\nu}
\bigl(\nabla_\rho b\,\nabla^\rho b + \mu_b^2 b^2\bigr).
\label{eq:Tmunu_axion}
\end{equation}
When the CS term is present, its metric variation introduces a Cotton-tensor
correction $C_{\mu\nu}$~\cite{Jackiw:2003pm}, giving the equation of motion  $G_{\mu\nu} + C_{\mu\nu}
= \kappa^2 T_{\mu\nu}^{(b)}$ with 
\begin{equation}\label{eq:Cotton}
C^{\mu\nu}
=
-(\nabla_\rho b)\,
\epsilon^{\rho\sigma\alpha(\mu}
\nabla_\alpha R^{\nu)}{}_\sigma
-
(\nabla_\rho\nabla_\sigma b)\,
{}^{*}R^{\sigma(\mu\nu)\rho}.
\end{equation}
and $G_{\mu\nu} = R_{\mu\nu} - \frac{1}{2}\, g_{\mu\nu}\, R$ the Einstein tensor. Although the Chern--Simons contribution is often described at the level
of the field equations through the Cotton tensor $C_{\mu\nu}$, the
quantity entering the graviton squeezing kernel is the corresponding
quadratic graviton vertex. Equivalently, it is obtained from the second
variation of the action term $b\,{}^{*}RR$, or from the first variation
of the Cotton tensor about the chosen background.
For the Chern--Simons channel it is not necessary to calculate the
Cotton tensor explicitly. For the flat-space approximation used here,
the background Riemann tensor vanishes,
\[
R_{\mu\nu\rho\sigma}[\eta]=0,
\]
so that the curvature expansion begins at first order:
\[
R_{\mu\nu\rho\sigma}
=
R^{(1)}_{\mu\nu\rho\sigma}[h]
+
R^{(2)}_{\mu\nu\rho\sigma}[h,h]
+\cdots.
\]
For a Kerr background, one has instead schematically
\[
R_{\mu\nu\rho\sigma}
=
\bar R_{\mu\nu\rho\sigma}
+
R^{(1)}_{\mu\nu\rho\sigma}[h]
+
R^{(2)}_{\mu\nu\rho\sigma}[h,h]
+\cdots,
\]
(and additional terms involving the background curvature would contribute
to the Chern--Simons dynamics).
Explicitly 
\begin{equation}
{
R^{(1)}_{\mu\nu\rho\sigma}[h]
=
\frac{\kappa}{2}
\Big(
\partial_\rho\partial_\nu h_{\mu\sigma}
+
\partial_\sigma\partial_\mu h_{\nu\rho}
-
\partial_\rho\partial_\mu h_{\nu\sigma}
-
\partial_\sigma\partial_\nu h_{\mu\rho}
\Big).
}
\label{eq:R1_explicit}
\end{equation}
and, on substituting the explicit Christoffel symbols, the contributions to $R^{(2)}$ yield terms such as
\begin{equation}
R^{(2)}
\sim
\kappa^2
\left[
h\,\partial\partial h
+
(\partial h)(\partial h)
\right].
\end{equation}
Therefore the Pontryagin density begins at quadratic order,
\begin{equation}
{}^{*}RR
=
{}^{*}R^{(1)}_{\mu\nu\rho\sigma}[h]\,
R^{(1)\mu\nu\rho\sigma}[h]
+
\mathcal O(h^3).
\end{equation}

Consequently the CS interaction
\begin{equation}
S_{\rm CS}
=
-\int d^4x\,
A\,b_{\rm cl}(x)\,{}^{*}RR
\end{equation}
already contains a direct two-graviton vertex:
\begin{equation}
S_{\rm CS}^{(2)}
=
-\int d^4x\,
A\,b_{\rm cl}(x)\,
{}^{*}R^{(1)}_{\mu\nu\rho\sigma}[h]\,
R^{(1)\mu\nu\rho\sigma}[h].
\end{equation}

This quadratic term is the contribution that enters the graviton
squeezing kernel. Omitting details,
\begin{equation}
Z^{\rm CS}_{ij}
\sim
-iA
\int d^4x\,
b_{\rm cl}(x)\,
{}^{*}R^{(1)}[u_i]\,
R^{(1)}[u_j],
\end{equation}
where $u_i$ and $u_j$ denote the two outgoing graviton mode functions \cite{Dorlis:2025amf,Dorlis:2025zzz}.

To see how the Chern--Simons interaction modifies the squeezing kernel,
we expand the graviton perturbation in transverse--traceless modes,
\begin{equation}
h_{\mu\nu}(x)
=
\sum_{h=\pm2}\int \frac{d^3k}{(2\pi)^3}
\left[
\epsilon^{(h)}_{\mu\nu}(\hat{\mathbf k})
u_{\mathbf k}(x)\,
a_h(\mathbf k)
+
\epsilon^{(h)*}_{\mu\nu}(\hat{\mathbf k})
u_{\mathbf k}^*(x)\,
a_h^\dagger(\mathbf k)
\right].
\label{eq:graviton_mode_expansion}
\end{equation}
Here \(h=\pm2\) labels the two circular graviton helicities,
\(\mathbf k\) is the spatial momentum, and \(u_{\mathbf k}(x)\) is the
corresponding mode function.

The polarisation tensors
\(\epsilon^{(h)}_{\mu\nu}(\hat{\mathbf k})\) are symmetric,
transverse and traceless:
\begin{equation}
\epsilon^{(h)}_{\mu\nu}
=
\epsilon^{(h)}_{\nu\mu},
\qquad
k^\mu\epsilon^{(h)}_{\mu\nu}=0,
\qquad
\eta^{\mu\nu}\epsilon^{(h)}_{\mu\nu}=0.
\end{equation}
In a purely spatial transverse--traceless gauge one may take
\begin{equation}
\epsilon^{(h)}_{0\mu}=0,
\qquad
k^i\epsilon^{(h)}_{ij}=0,
\qquad
\delta^{ij}\epsilon^{(h)}_{ij}=0.
\end{equation}
The tensors may be built from circular vector polarisations
\(e_i^{(\pm)}(\hat{\mathbf k})\) as
\begin{equation}
\epsilon^{(+2)}_{ij}(\hat{\mathbf k})
=
e_i^{(+)}(\hat{\mathbf k})
e_j^{(+)}(\hat{\mathbf k}),
\qquad
\epsilon^{(-2)}_{ij}(\hat{\mathbf k})
=
e_i^{(-)}(\hat{\mathbf k})
e_j^{(-)}(\hat{\mathbf k}).
\end{equation}
The vector polarisations obey
\begin{equation}
\hat k^i e_i^{(\pm)}=0,
\qquad
e_i^{(\pm)*}e_i^{(\pm)}=1,
\qquad
e_i^{(+)*}e_i^{(-)}=0.
\end{equation}
With this convention the graviton helicity \(h=+2\) corresponds to the
right-handed circular tensor polarisation and \(h=-2\) to the left-handed
one, up to the chosen helicity convention.

Equivalently, one may combine the helicity and momentum labels into a
single composite-index
\begin{equation}
i\equiv(h,\mathbf k),
\end{equation}
so that
\begin{equation}
h_{\mu\nu}(x)
=
\sum_i
\left[
\epsilon^{(i)}_{\mu\nu}
u_i(x)a_i
+
\epsilon^{(i)*}_{\mu\nu}
u_i^*(x)a_i^\dagger
\right].
\end{equation}
\color{black} The Chern--Simons interaction contains, at quadratic order in the
graviton perturbation,
\begin{equation}
S_{\rm CS}^{(2)}
=
-\int d^4x\,
A\,b_{\rm cl}(x)\,
{}^{*}R^{(1)}_{\mu\nu\rho\sigma}[h]\,
R^{(1)\mu\nu\rho\sigma}[h].
\end{equation}
Substituting the mode expansion of $h_{\mu\nu}$ into this expression
produces terms proportional to
\begin{equation}
a_i^\dagger a_j^\dagger ,
\end{equation}
which create correlated graviton pairs \cite{Dorlis:2025amf,Dorlis:2025zzz}. The coefficient of this pair
creation operator defines the Chern--Simons contribution to the
squeezing kernel:
\begin{equation}
Z^{\rm CS}_{ij}
\sim
-iA
\int d^4x\,
b_{\rm cl}(x)\,
{}^{*}R^{(1)}[u_i^*]\,
R^{(1)}[u_j^*].
\end{equation}
Equivalently,
\begin{equation}
b\,{}^{*}RR
\longrightarrow
b_{\rm cl}\,
{}^{*}R^{(1)}[h]R^{(1)}[h]
\longrightarrow
\text{graviton pair creation}
\longrightarrow
Z^{\rm CS}_{ij}.
\end{equation}
The axion cloud is populated to exponentially large occupation numbers by
superradiance. Its expectation value is approximated by the classical bound-state
profile
\begin{equation}
b_{\rm cl}(t,\mathbf{x})
\simeq
b_0\,
R_{nlm}(r)\,
Y_{lm}(\theta,\phi)\,
\cos\!\left(
\omega_b t
-
m\phi
+
\delta
\right),
\end{equation}
where \(b_0\) is the cloud amplitude, \(R_{nlm}(r)\) is the radial
bound-state wavefunction, \(Y_{lm}\) is the angular harmonic,
\(\omega_b\) is the axion bound-state frequency, and \(\delta\) is an
overall phase. The bound-state frequency $\omega_b \simeq \mu_b(1 - \alpha_g^2/2s^2)$ and
gravitational fine-structure constant $\alpha_g \equiv GM_{\rm BH}\mu_b$. Relative quantum fluctuations of the cloud are suppressed as $\Delta N/N \sim
N^{-1/2} \ll 1$, justifying the classical treatment. For details of $b_{cl}$ see Appendix \ref{app:axion_cloud_profile}.

\subsection{Origin of the squeezed state in the rotating wave approximation}

The \emph{dominant} gravitational channel in the axion cloud is coherent fusion of two axion quanta into
a graviton pair (with frequencies $\Omega_1$ and $\Omega_2$),  $b + b \to g + g$, generating a pair-creation term
$a^\dagger_{\mathbf{k}_1} a^\dagger_{\mathbf{k}_2}
e^{-i(\Omega_1+\Omega_2-2\omega_b)t}$ in the interaction Hamiltonian \cite{Dorlis:2025amf,Dorlis:2025zzz}.
Retaining only near-resonant terms is the rotating wave
approximation (RWA)~\cite{ScullyZubairy,CohenTannoudji}) that imposes the condition
\begin{equation}
\Omega_1 + \Omega_2 \simeq 2\omega_b.
\end{equation}
A subleading Chern-Simons channel ($b\to g+g$) gives $\Omega_1+\Omega_2 \simeq
\omega_b$; this channel is quantitatively suppressed, but  
introduces qualitatively new \emph{helicity asymmetries}~\cite{Dorlis:2025zzz,Dorlis:2025amf}. RWA is justified because the relevant
frequency scales are well separated. The axion cloud is non-relativistic,
so its bound-state frequencies are sharply peaked around the axion mass,
with a small spread of order
\[
\Delta\Omega \sim \alpha_g^2 \mu_a .
\]
The graviton-pair source therefore has a narrow bandwidth. Terms in the
interaction Hamiltonian, whose phases oscillate much faster than this
bandwidth, average to zero over the timescale on which the cloud evolves.
Only terms satisfying the approximate resonance condition are retained.

 In addition, the interaction is suppressed by the Planck scale through
the universal gravitational coupling
\[
{\cal L}_{\rm int}
\sim
\frac{1}{M_{\rm Pl}}
h_{\mu\nu}T^{\mu\nu}.
\]
The corresponding dimensionless expansion parameter is therefore of
order
\[
\frac{\Omega}{M_{\rm Pl}},
\]
where \(\Omega\) denotes the characteristic graviton frequency.
For the axion-cloud systems considered here this quantity is extremely
small, ensuring that the interaction changes the meson state only on
timescales much longer than the microscopic oscillation periods. Finally, the axion cloud
itself evolves adiabatically compared with the graviton oscillation time.
These scale separations make the rotating-wave approximation
parametrically controlled: one keeps the slowly varying near-resonant
terms and discards rapidly oscillating counter-rotating terms.

The resonant pair-production process yields a graviton state of the form \cite{Dorlis:2025amf,Dorlis:2025zzz}
\begin{equation}
|\Psi_{\text{grav}}\rangle
=
\exp\!\left[
\frac{1}{2}\sum_{ij}
\bigl(\mathcal{G}_{ij} a_i^\dagger a_j^\dagger - \mathcal{G}_{ij}^* a_i a_j\bigr)
\right]|0\rangle,
\label{eq:squeezed_state_general}
\end{equation}
where \footnote{Here we have used $a$ to denote the graviton annihilation operator rather than $\alpha$, which was the corresponding notation in~\cite{Dorlis:2025amf}. }  $\mathcal{G}_{ij} = \mathcal{G}_{ji}$ is a
complex \emph{symmetric} squeezing kernel. The kernel is obtained by projecting the
quadratic axion source onto pairs of outgoing graviton modes (see \cite{Dorlis:2025zzz,Dorlis:2025amf} and
Appendix~\ref{app:Jkernel} for a derivation); it is not diagonal in general,
reflecting correlated momenta and helicities. The non-diagonal structure of \(\mathcal G_{ij}\) implies that the
squeezing is generally distributed across many graviton modes and that
different momentum and helicity channels are quantum mechanically
correlated. Nevertheless, because \(\mathcal G_{ij}\) is complex
symmetric, it admits a Takagi decomposition \cite{Houde:2024mkj} through a unitary operator $U$, which
defines a new set of graviton operators,
\begin{equation}
\label{TakagiSupermode}
\mathbb b_\alpha
=
\sum_i
U^*_{i\alpha}a_i,
\end{equation}
for which the multimode squeezed state factorises into independent
squeezed supermodes. In this basis the mode correlations are absorbed
into the definition of the supermodes themselves, and the squeezing
operator becomes a product of commuting single-mode squeezing operators.
The resulting supermodes provide the natural basis for analysing the
open-system dynamics of the neutral-meson sector.

\color{black}\paragraph{Effect of the gravitational Chern--Simons interaction.}
The  gravitational Chern--Simons term  produces a
perturbative deformation
\[
\mathcal{G}_{ij}^{\rm GR}
\;\longrightarrow\;
\mathcal{G}_{ij}^{\rm GR+CS}
=
\mathcal{G}_{ij}^{\rm GR}
+
\delta\mathcal{G}_{ij}^{\rm CS}.
\]
This deformation modifies the Takagi spectrum (see Appendix \ref{app:takagi})
$(r_\alpha,\phi_\alpha,U_{i\alpha})\to(R_\alpha,\Phi_\alpha,\widetilde{U}_{i\alpha})$
and therefore changes the magnitude and phase of the resulting
$\omega$-parameter. The Chern--Simons interaction thus affects the
$\omega$-effect \emph{indirectly}, through its modification of the
squeezing kernel.
In pure GR, graviton pairs are predominantly produced with
opposite helicities and momenta, so that 
\begin{align}\label{ZGR}
Z^{\rm GR}_{RL}=Z^{\rm GR}_{LR}\,,
\end{align}
whereas the parity-breaking Chern--Simons term adds a
helicity-antisymmetric contribution,
\begin{equation}
Z_{hh'}^{\rm GR+CS} = Z_{hh'}^{\rm GR} + Z_{hh'}^{\rm CS},
\qquad Z^{\rm CS}_{RL} = -Z^{\rm CS}_{LR}.
\label{eq:TakagiCS}
\end{equation}
The polarization structure of the squeezed graviton state is also modified
by the Chern--Simons interaction. In the pure-GR calculation of
Refs.~\cite{Dorlis:2025zzz,Dorlis:2025amf,Dorlis:2026gth}, the resulting
squeezing kernel was found to treat the two circular graviton helicities
symmetrically, in the sense that the \(L\) and \(R\) sectors enter with the
same weight after the mode sums and angular projections are performed.
The Chern-Simons term changes this result. Being parity odd, it weights
the two helicities differently and therefore produces a chiral deformation
of the squeezing kernel.

The difference between the pure-GR and Chern--Simons contributions should
be interpreted carefully. In the graviton sector the Chern--Simons term
introduces an explicit parity-violating interaction in the microscopic
action. It therefore modifies the squeezing kernel by weighting the two
circular graviton helicities differently. This is a consequence of the
parity-odd structure of \(b\,{}^\star RR\), not an emergent effect caused
by tracing over an environment.

This should be distinguished from the neutral-meson \(\omega\)-effect.
There, the microscopic dynamics remains CPT invariant, while CPT
becomes ill-defined only for the reduced meson subsystem after the
gravitational environment has been traced out \cite{Wald:1980}. Thus the CS deformation is
an explicit parity-violating modification of the gravitational bath,
whereas the meson \(\omega\)-effect is an open-system effect in the
neutral-kaon sector.
\color{black}
\subsection{Takagi decomposition and supermode structure}
\label{sec:Takagi}

Since $\mathcal{G}_{ij}$ is complex symmetric, it admits a Takagi
factorisation~\cite{Takagi:1925,Houde:2024mkj} which has the form,
\begin{equation}
\mathcal{G} = U R U^T,
\qquad R = \mathrm{diag}(r_1,r_2,\dots),\quad r_\alpha \ge 0,
\end{equation}
with $U$ unitary (see Appendix \ref{app:takagi}.). 
From calculations \cite{Dorlis:2025amf,Dorlis:2025amf} for axion clouds,  we find
\begin{equation}
\mathcal G^{\rm GR}_{(R,\mathbf{k})(L,\mathbf{k}')}
\simeq
\mathcal G^{\rm GR}_{(L,\mathbf{k})(R,\mathbf{k}')},
\end{equation}
and 
\begin{equation}
\mathcal G^{\rm CS}_{(R,\mathbf{k})(L,\mathbf{k}')}
\simeq
-
\mathcal G^{\rm CS}_{(L,\mathbf{k})(R,\mathbf{k}')}.
\end{equation}
The full squeezing kernel contains both equal-helicity and
opposite-helicity sectors,
\begin{equation}
\mathcal G
=
\begin{pmatrix}
\mathcal G_{RR} & \mathcal G_{RL}\\
\mathcal G_{LR} & \mathcal G_{LL}
\end{pmatrix},
\end{equation}
where each entry is itself a momentum-space kernel
\(\mathcal G_{hh'}(\mathbf k,\mathbf k')\).
In the dominant pure-GR axion-cloud channel one usually expects the
opposite-helicity blocks to dominate,
\begin{equation}
\mathcal G^{\rm GR}_{RL},
\,
\mathcal G^{\rm GR}_{LR}
\gg
\mathcal G^{\rm GR}_{RR},
\,
\mathcal G^{\rm GR}_{LL}.
\end{equation}
In the CS-deformed case the dominant helicity correlations are expected
to remain in the opposite-helicity sector, but with a different chirality.

We use Takagi supermodes
$\mathbb{b}_\alpha = \sum_i U^*_{i\alpha} a_i$ \eqref{TakagiSupermode} ,
which satisfy canonical commutation relations owing to the unitarity of $U$,
and on expanding the graviton field in Takagi supermodes (see Appendix \ref{app:takagi}),
\begin{equation}
\label{TakagiExpansion}
h_{\mu\nu}(x)
=
\sum_\alpha
\Bigl[
{\cal U}^{(\alpha)}_{\mu\nu}(x)\,
\mathbb b_\alpha
+
{\cal U}^{(\alpha)*}_{\mu\nu}(x)\,
\mathbb b_\alpha^\dagger
\Bigr],
\end{equation}
the graviton state \cite{Dorlis:2025amf}
factorises into independent single-mode squeezers:
\begin{equation}
|\Psi_{\text{grav}}\rangle
=
\prod_\alpha
\exp\!\left[
\tfrac{r_\alpha}{2}
\bigl(e^{-i\phi_\alpha} \mathbb{b}_\alpha^2
- e^{i\phi_\alpha} \mathbb{b}_\alpha^{\dagger 2}\bigr)
\right]
|0_\alpha\rangle=
\prod_\alpha
S_\alpha(r_\alpha,\phi_\alpha)
|0_\alpha\rangle ..
\end{equation}
The supermode correlators are
\begin{align}
\langle \mathbb{b}_\alpha^\dagger \mathbb{b}_\alpha \rangle
&= \sinh^2 r_\alpha,
\label{eq:normal_corr}\\
\langle \mathbb{b}_\alpha \mathbb{b}_\alpha \rangle
&= -e^{i\phi_\alpha}\sinh r_\alpha \cosh r_\alpha.
\label{eq:anomalous_corr}
\end{align}
The normal correlator describes stimulated processes present for any non-vacuum
bath. The \emph{anomalous} correlator~\eqref{eq:anomalous_corr} is the
signature of pair production; it vanishes identically for a thermal, coherent,
or classical gravitational noise and is precisely the quantity that drives the
$\omega$-effect, as shown in Section~\ref{sec:interaction}.It is important to distinguish a two-graviton state from a squeezed
vacuum. A single pair-production event produces a state proportional to
\[
a_i^\dagger a_j^\dagger|0\rangle .
\]
This is not, by itself, a squeezed state. A squeezed vacuum arises when
     the axion cloud acts as a coherent, approximately classical pump that
drives repeated pair creation. In that case the time-evolution operator
generated by a quadratic pair-production Hamiltonian exponentiates:
\[
U
=
\exp\left[
\frac12
\sum_{ij}
\left(
{\cal G}_{ij}a_i^\dagger a_j^\dagger
-
{\cal G}_{ij}^*a_i a_j
\right)
\right],
\]
and acting on the reference vacuum gives
\[
|\Psi_{\rm grav}\rangle=U|0\rangle .
\]
The two-graviton amplitude is then the leading term in the expansion of
this multimode squeezed state, while higher even-graviton components are
generated by repeated coherent pair production.

\subsection{From stochastic fluctuations to quantum graviton correlations}
\label{sec:stochastic_dict}

It is instructive to compare the present derivation with the stochastic
D-foam model reviewed in the Introduction. There the $\omega$-effect was
controlled by the variance $\Delta_i = \langle r_i^2\rangle$ of
phenomenological random variables encoding the strength of spacetime
defect scattering.  \cbl{As mentioned there, it is the small mass difference of the (near degenerate) neutral-meson mass eigenstates that acts as a ``magnifying'' factor that compensates for the quadratic-Planck-energy scale sup[pression of the stochastic variance $\Delta_2$, which is an expected feature of generic quantum-gravity environments.}

In the present quantum treatment there are no stochastic variables
$r_i$. The gravitational environment is described by the pure state
$\rho_{\rm grav} = |\Psi_{\rm grav}\rangle\langle\Psi_{\rm grav}|$, and
stochastic averages are replaced by expectation values in this squeezed
graviton state. The crucial quantity is the \emph{anomalous} correlator
\begin{equation}
\langle \mathbb{b}_\alpha \mathbb{b}_\beta\rangle
=
-\delta_{\alpha\beta}\,e^{i\phi_\alpha}\sinh r_\alpha\cosh r_\alpha ,
\end{equation}
which replaces the D-foam variance:
\begin{equation}
\Delta_i = \langle r_i^2\rangle
\quad\longrightarrow\quad
e^{i\phi_\alpha}\sinh r_\alpha\cosh r_\alpha .
\label{eq:dictionary}
\end{equation}
The replacement~\eqref{eq:dictionary} is not merely formal. The anomalous
correlator $\langle\mathbb{b}_\alpha\mathbb{b}_\alpha\rangle$ vanishes
identically for any thermal, coherent, or classically noisy gravitational
bath and is a unique signature of pair production in the squeezed state.
By contrast, the normal correlator $\langle\mathbb{b}_\alpha^\dagger
\mathbb{b}_\alpha\rangle = \sinh^2 r_\alpha$ is present for any non-vacuum
bath. As shown in Section~\ref{sec:interaction}, it is the anomalous
correlator that generates the simultaneous two-meson operators
$\sigma_+^{(1)}\sigma_+^{(2)}$ and $\sigma_-^{(1)}\sigma_-^{(2)}$
responsible for mixing the antisymmetric and symmetric sectors; normal
correlators contribute only to ordinary decoherence and dispersive
shifts. This is why a squeezed graviton bath is \emph{necessary} for
the $\omega$-effect: a thermal or classical gravitational noise cannot
generate it.

The stochastic D-foam model also treated the environment as
time-independent, so no dynamical time structure entered the estimate
for $\omega$. In the present treatment the environment has resolved
quantum modes with frequencies $\Omega_\alpha$, and the response of the
meson system depends on the detuning $\Omega_\alpha - \Delta E$ and on
the duration of the interaction. This dynamical information is encoded
in the time kernel $F_\alpha(t)$ derived in Section~\ref{sec:omega}.

\section{Meson--Graviton Interaction and Reduced Meson Dynamics}
\label{sec:interaction}

\subsection{The meson Hamiltonian and interaction}

We describe each neutral meson as a two-level system. \cbl{For concreteness 
we shall use the neutral kaon notation, although our approach is actually generic and applies to all types of entangled neutral mesons.}

Before discussing the meson--graviton interaction, it is useful to
distinguish the three Hamiltonians that enter the analysis.

The starting point is the standard neutral-kaon effective Hamiltonian,
defined in the flavour basis
\(\{|K^0\rangle,|\bar K^0\rangle\}\),
\begin{equation}
H_K
=
M-\frac{i}{2}\Gamma,
\end{equation}
where the off-diagonal entries
\(M_{12}-i\Gamma_{12}/2\) and
\(M_{21}-i\Gamma_{21}/2\)
are generated by weak interactions through
\(|\Delta S|=2\) processes.
These weak-interaction terms cause
\(K^0\) and \(\bar K^0\) to oscillate into one another.

Diagonalising \(H_K\) yields the physical propagation eigenstates\footnote{Beyond the CP-conserving
approximation, the mixing coefficients are
\begin{equation}
p = \frac{1+\varepsilon}{\sqrt{2(1+|\varepsilon|^2)}},
\qquad
q = \frac{1-\varepsilon}{\sqrt{2(1+|\varepsilon|^2)}},
\end{equation}
where $\varepsilon$ is the CP-violation parameter, $|\varepsilon|\approx
2.2\times 10^{-3}$ for the case of neutral Kaons~\cite{ParticleDataGroup:2024cfk}.}
\begin{equation}
|K_S\rangle
=
p|K^0\rangle+q|\bar K^0\rangle,
\qquad
|K_L\rangle
=
p|K^0\rangle-q|\bar K^0\rangle,
\end{equation}
with complex eigenvalues
\begin{equation}
\lambda_{S,L}
=
m_{S,L}
-\frac{i}{2}\Gamma_{S,L}.
\end{equation}

Since the weak-interaction mixing has already been incorporated into the
definition of \(K_S\) and \(K_L\), the kaon subsystem may be represented as
an effective two-level system with free Hamiltonian
\begin{equation}
H_M
=
\frac{\Delta\lambda}{2}\sigma_3,
\qquad
\Delta\lambda
=
\lambda_L-\lambda_S,
\end{equation}
where
\begin{equation}
\sigma_3
=
|K_L\rangle\langle K_L|
-
|K_S\rangle\langle K_S|.
\end{equation}

The second ingredient is the interaction between this two-level system
and the squeezed graviton environment. The total Hamiltonian therefore
takes the form
\begin{equation}
H
=
H_M
+
H_{\rm grav}
+
H_{\rm int},
\end{equation}
where \(H_{\rm grav}\) describes the Takagi supermodes of the squeezed
graviton bath and \(H_{\rm int}\) couples the mesons to those modes (and also has a contribution from the weak interactions).
Only after tracing over the graviton degrees of freedom will we obtain a
non-unitary evolution equation for the reduced meson density matrix.
The reduced-density-matrix description therefore appears at a later stage;
at the level of the microscopic theory the dynamics is generated by the
ordinary Hamiltonian \(H\).

The effective meson--graviton interaction originates from the universal
coupling of gravity to the meson energy--momentum tensor (\emph{cf.} \eqref{tmnmeson}),
\begin{equation}
S_{\rm int}
=
\frac12
\int d^4x\,
h_{\mu\nu}(x)\,
T^{\mu\nu}(x).
\label{eq:Sint_Tmunu}
\end{equation}

After expanding the graviton field in Takagi supermodes,
the interaction takes the form
\begin{equation}\label{Hint1}
H_{\rm int}
=
\sum_\alpha
\Bigl(
{\cal T}_\alpha\,\mathbb b_\alpha
+
{\cal T}_\alpha^\dagger\,\mathbb b_\alpha^\dagger
\Bigr),
\end{equation}
where
\begin{equation}\label{Hint2}
{\cal T}_\alpha
=
\frac12
\int d^3x\,
{\cal U}^{(\alpha)}_{\mu\nu}(x)
T^{\mu\nu}(x).
\end{equation}

The neutral-kaon sector is treated as an effective two-level system
spanned by the propagation eigenstates
\(
|K_S\rangle
\)
and
\(
|K_L\rangle.
\)
The operator obtained by contracting the graviton mode with the
energy--momentum tensor is projected onto this two-dimensional subspace,
\[
X_\alpha
=
P_{KL,KS}\,
{\cal T}_\alpha\,
P_{KL,KS},
\]
where
\[
P_{KL,KS}
=
|K_S\rangle\langle K_S|
+
|K_L\rangle\langle K_L|.
\]
The resulting operator acts only on the effective two-level kaon system
and may therefore be expanded in the Pauli basis. This is elaborated on in Appendix \ref{app:Tmunu}


To describe the neutral-kaon system as an effective two-level system, one
projects the operator ${\cal T}_\alpha$ onto the subspace spanned by
\(\{|K_S\rangle,|K_L\rangle\}\). The resulting projected operator is
denoted by \(X\),
\begin{equation}
X
=
P_{KL,KS}\,
{\cal T}_\alpha\,
P_{KL,KS},
\end{equation}
where \(P_{KL,KS}\) is the projector onto the
\(\{|K_S\rangle,|K_L\rangle\}\) subspace.
Since any operator acting on a two-dimensional Hilbert space can be
expanded in the Pauli basis, one may write
\begin{equation}
X
=
c_0{\bf 1}
+
c_3\sigma_3
+
c_+\sigma_+
+
c_-\sigma_- .
\end{equation}
The gravitational coupling of the meson pair to the squeezed bath is
\begin{equation}
H_{\text{int}}
=
\sum_{a=1}^{2}\sum_\alpha
\Bigl(g_\alpha^{(a)} X^{(a)} \mathbb{b}_\alpha
+ g_\alpha^{(a)*} X^{(a)} \mathbb{b}_\alpha^\dagger\Bigr),
\label{eq:Hint}
\end{equation}
where $a=1,2$ labels the two mesons, $g_\alpha^{(a)}$ are the effective
couplings to the $\alpha$-th graviton supermode, and $X^{(a)}$ is the
projection of the energy--momentum tensor $T_{\mu\nu}$ onto the two-level meson
Hilbert space. Any such operator decomposes as $X = c_0\mathbf{1} + c_3\sigma_3
+ c_+\sigma_+ + c_-\sigma_-$, where
\begin{equation}
\sigma_+ = |K_L\rangle\langle K_S|, \qquad \sigma_- = |K_S\rangle\langle K_L|.
\end{equation}
The terms proportional to $\mathbf{1}$ and $\sigma_3$ preserve the
$K_S$--$K_L$ quantum numbers and do not mix flavour sectors. Only the
transition part
\begin{equation}
X_{\rm tr} = c_+\sigma_+ + c_-\sigma_-
\end{equation}
 contributes to the $\omega$-effect where 
\begin{equation}
{
c_+
=
c_1-i c_2,
\qquad
c_-
=
c_1+i c_2.
}
\end{equation}
Crucially, non-zero coefficients $c_\pm$
require off-diagonal matrix elements $\langle K_S|X|K_L\rangle \neq 0$. In the
free mass-eigenstate theory the stress tensor is diagonal in the $K_S$--$K_L$
basis, giving $c_\pm = 0$. It is precisely the weak-interaction mixing of
$K^0$ and $\bar{K}^0$ that generates off-diagonal components when the
microscopic stress tensor is projected onto the propagation-eigenstate basis:
since $K_S$ and $K_L$ are each superpositions of $K^0$ and $\bar{K}^0$, any
operator that is off-diagonal in the flavour basis (such as the strangeness-
changing part of the effective weak Hamiltonian) acquires non-zero transition
matrix elements between $K_S$ and $K_L$. 

The total density matrix obeys
\begin{equation}
\dot\rho_{\rm tot}
=
-i[H,\rho_{\rm tot}],
\end{equation}
where
\begin{equation}
H
=
H_M
+
H_{\rm grav}
+
H_{\rm int}.
\end{equation}
We now work in the interaction picture with respect to
\begin{equation}
H_0
=
H_M+H_{\rm grav},
\end{equation}
where the free graviton Hamiltonian is diagonal in the Takagi basis,
\begin{equation}
H_{\rm grav}
=
\sum_\alpha
\Omega_\alpha
\mathbb b_\alpha^\dagger
\mathbb b_\alpha .
\end{equation}
Thus the free interaction-picture evolution of the graviton supermodes is
\begin{equation}
\mathbb b_\alpha(t)
=
e^{iH_{\rm grav}t}
\mathbb b_\alpha
e^{-iH_{\rm grav}t}
=
\mathbb b_\alpha e^{-i\Omega_\alpha t},
\end{equation}
and
\begin{equation}
\mathbb b_\alpha^\dagger(t)
=
\mathbb b_\alpha^\dagger e^{+i\Omega_\alpha t}.
\end{equation}
Similarly, for the meson transition operators,
\begin{equation}
X(t)
=
e^{iH_Mt}Xe^{-iH_Mt}.
\end{equation}
If
\begin{equation}
H_M
=
\frac{\Delta\lambda}{2}\sigma_3,
\end{equation}
then
\begin{equation}
\sigma_+(t)
=
\sigma_+ e^{+i\Delta\lambda t},
\qquad
\sigma_-(t)
=
\sigma_- e^{-i\Delta\lambda t}.
\end{equation}
Therefore
\begin{equation}
\label{mesonInteraction}
X(t)
=
c_+\sigma_+ e^{+i\Delta\lambda t}
+
c_-\sigma_- e^{-i\Delta\lambda t}
+\cdots 
\end{equation}
where the terms proportional to \(\mathbf 1\) and \(\sigma_3\) have been
suppressed because they do not contribute to the \(\omega\)-effect. Working in the interaction picture 
the total density matrix obeys
\[
\frac{d\tilde\rho_{\rm tot}}{dt}
=
-i[\tilde H_{\rm int}(t),\tilde\rho_{\rm tot}(t)].
\]
and
\begin{equation}
\langle
\mathbb b_\alpha(t_1)
\mathbb b_\beta(t_2)
\rangle
=
-\delta_{\alpha\beta}
e^{i\phi_\alpha}
\sinh r_\alpha\cosh r_\alpha
e^{-i\Omega_\alpha(t_1+t_2)} .
\end{equation}
Assuming weak meson--graviton coupling and using the Born approximation,
\[
\tilde\rho_{\rm tot}(s)
\simeq
\tilde\rho_M(s)\otimes\rho_{\rm grav},
\]
one obtains
\[
\frac{d\tilde\rho_M}{dt}
=
-\int_0^t ds\,
{\rm Tr}_{\rm grav}
\left[
\tilde H_{\rm int}(t),
[
\tilde H_{\rm int}(s),
\tilde\rho_M(s)\otimes\rho_{\rm grav}
]
\right].
\]
Assuming weak meson--graviton coupling, we trace over the graviton
environment to yield the second-order master equation
\begin{equation}
\label{SecondOrderMaster}
\dot{\tilde{\rho}}_M(t)
=
-
\int_0^t ds\,
{\rm Tr}_{\rm grav}
\Big[
\tilde H_{\rm int}(t),
\big[
\tilde H_{\rm int}(s),
\tilde\rho_M(s)\otimes\rho_{\rm grav}
\big]
\Big].
\end{equation}
Substituting Eq.~(\ref{eq:Hint}) into
Eq.~(\ref{SecondOrderMaster}) generates products of two graviton operators.
Tracing over the bath produces two classes of interaction-picture
two-point functions. The normal correlators are
\begin{equation}
\left\langle
\mathbb b_\alpha^\dagger(t_1)
\mathbb b_\beta(t_2)
\right\rangle
=
\delta_{\alpha\beta}
\sinh^2 r_\alpha\,
e^{+i\Omega_\alpha t_1}
e^{-i\Omega_\alpha t_2},
\end{equation}
and
\begin{equation}
\left\langle
\mathbb b_\alpha(t_1)
\mathbb b_\beta^\dagger(t_2)
\right\rangle
=
\delta_{\alpha\beta}
\left(\sinh^2 r_\alpha+1\right)
e^{-i\Omega_\alpha t_1}
e^{+i\Omega_\alpha t_2}.
\end{equation}
These describe ordinary absorption, emission, and stimulated processes.

The anomalous correlators are
\begin{equation}
\left\langle
\mathbb b_\alpha(t_1)
\mathbb b_\beta(t_2)
\right\rangle
=
-\delta_{\alpha\beta}
e^{i\phi_\alpha}
\sinh r_\alpha\cosh r_\alpha\,
e^{-i\Omega_\alpha(t_1+t_2)},
\end{equation}
and
\begin{equation}
\left\langle
\mathbb b_\alpha^\dagger(t_1)
\mathbb b_\beta^\dagger(t_2)
\right\rangle
=
-\delta_{\alpha\beta}
e^{-i\phi_\alpha}
\sinh r_\alpha\cosh r_\alpha\,
e^{+i\Omega_\alpha(t_1+t_2)}.
\end{equation}
The anomalous correlators are non-zero only for a squeezed state.
They are the terms that multiply products such as
\(X^{(1)}(t_1)X^{(2)}(t_2)\) in the Born kernel and generate the
simultaneous two-meson transitions relevant for the \(\omega\)-effect.
The anomalous bath contraction multiplies products of interaction-picture
meson operators of the form
\begin{equation}
X^{(1)}(t_1)X^{(2)}(t_2).
\end{equation}
Passing to the interaction picture with respect to the free meson
Hamiltonian \eqref{mesonInteraction}
gives

\begin{align}
X^{(1)}(t_1)X^{(2)}(t_2)
=
&
\,c_+^2
e^{+i\Delta\lambda(t_1+t_2)}
\sigma_+^{(1)}\sigma_+^{(2)}
\nonumber\\
&
+
c_-^2
e^{-i\Delta\lambda(t_1+t_2)}
\sigma_-^{(1)}\sigma_-^{(2)}
+\cdots .
\end{align}
The anomalous graviton correlator therefore multiplies precisely the
simultaneous two-meson transition operators
\begin{equation}
\sigma_+^{(1)}\sigma_+^{(2)},
\qquad
\sigma_-^{(1)}\sigma_-^{(2)},
\end{equation}
which connect the antisymmetric and symmetric sectors and hence generate
the \(\omega\)-effect.

The anomalous graviton correlators
$\langle \mathbb{b}_\alpha \mathbb{b}_\alpha\rangle$ thus generate the
$\omega$-effect only in conjunction with this weak-interaction-induced
flavour structure; a purely gravitational coupling to a diagonal stress tensor
would leave $\omega = 0$. 
The microscopic Hamiltonian~\eqref{eq:Hint} can remain CPT invariant;
CPT violation emerges only at the level of the reduced dynamics. The derivation
of $X$ from the meson stress tensor is given in Appendix~\ref{app:Tmunu}.

\section{Generation of the $\omega$ Parameter}
\label{sec:omega}

\subsection{Second-order transition amplitude and the main result}

The reduced meson density matrix is initially taken to be
\begin{equation}
\tilde\rho_M(0)
=
|A\rangle\langle A|.
\end{equation}
The quantity of interest is the matrix element
\begin{equation}
\omega(t)
\equiv
\langle S|
\tilde\rho_M(t)
|A\rangle,
\label{eq:omega_def_density}
\end{equation}
which measures the generation of a symmetric admixture in the reduced
meson state.

Substituting Eq.~(\ref{eq:Hint}) into the Born master equation
(\ref{SecondOrderMaster}) generates products of two bath operators.
After tracing over the graviton environment, the resulting kernel is
expressed in terms of bath two-point functions.
The contribution relevant for Eq.~(\ref{eq:omega_def_density})
comes from the anomalous correlator
\begin{equation}
\langle
\mathbb b_\alpha(t_1)
\mathbb b_\alpha(t_2)
\rangle
=
M_\alpha
e^{-i\Omega_\alpha(t_1+t_2)},
\end{equation}
with
\begin{equation}
M_\alpha
=
-
e^{i\phi_\alpha}
\sinh r_\alpha
\cosh r_\alpha .
\end{equation}
Projecting the master equation onto the
\(\langle S|\,\cdots\,|A\rangle\) matrix element yields
\begin{equation}
\omega(t)
=
-
\sum_\alpha
g_\alpha^{(1)}
g_\alpha^{(2)}
M_\alpha
\mathcal C_{AS}
\int_0^t dt_1
\int_0^{t_1}dt_2\,
e^{-i\delta_\alpha(t_1+t_2)}.
\end{equation}
\vskip .3cm
Starting from $|\psi(0)\rangle = |A\rangle$, the Born expansion
\eqref{SecondOrderMaster} induces an admixture of the symmetric sector. Since the
squeezed bath has vanishing one-point functions, the leading contribution
appears at second order. The relevant anomalous bath correlator is
\begin{equation}
\langle \mathbb{b}_\alpha(t_1)\mathbb{b}_\alpha(t_2)\rangle
= M_\alpha\,e^{-i\Omega_\alpha(t_1+t_2)},
\qquad
M_\alpha = -e^{i\phi_\alpha}\sinh r_\alpha\cosh r_\alpha,
\end{equation}
and the meson matrix element for the symmetry-violating transition is: 
\begin{equation}
\langle S|X^{(1)}(t_1)X^{(2)}(t_2)|A\rangle
= \mathcal{C}_{AS}\,e^{+i\Delta E(t_1+t_2)},
\end{equation}
where $\mathcal{C}_{AS}$ is the flavour-space matrix element (see
Appendix~\ref{app:Tmunu}). With detuning $\delta_\alpha \equiv
\Omega_\alpha-\Delta E$, the $\omega$-parameter is identified with the
transition amplitude $\omega(t) = \mathcal{A}_{A\to S}(t)$:
\begin{equation}
\omega(t)
=
-\sum_\alpha g_\alpha^{(1)}g_\alpha^{(2)}\,M_\alpha\,\mathcal{C}_{AS}
\int_0^t dt_1\int_0^{t_1}dt_2\;
e^{-i\delta_\alpha(t_1+t_2)}.
\end{equation}
Carrying out the integrals gives $\int_0^t dt_1\int_0^{t_1}dt_2\,
e^{-i\delta_\alpha(t_1+t_2)}
= -(1-e^{-i\delta_\alpha t})^2/(2\delta_\alpha^2)$,
so the main result is
\begin{equation}
\boxed{
\omega(t)
=
-\frac{1}{2}
\sum_\alpha
g_\alpha^{(1)}g_\alpha^{(2)}\,
e^{i\phi_\alpha}\sinh r_\alpha\cosh r_\alpha\;
\mathcal{C}_{AS}\;
\frac{\bigl(1-e^{-i(\Omega_\alpha-\Delta E)t}\bigr)^2}
{(\Omega_\alpha-\Delta E)^2}.
}
\label{eq:omega_main}
\end{equation}
This confirms that squeezing is both necessary and sufficient: all anomalous
correlators vanish for a thermal or coherent bath, giving $\omega\equiv 0$.
The magnitude and phase of $\omega$ are controlled by the product of couplings
$g_\alpha^{(1)}g_\alpha^{(2)}$, the squeezing data $(r_\alpha,\phi_\alpha)$ of
each Takagi supermode, the flavour matrix element $\mathcal{C}_{AS}$, and the
detuning $\Omega_\alpha-\Delta E$.

As noted earlier, including the gravitational Chern-Simons interaction does not introduce a
separate mechanism for generating the \(\omega\)-effect. Rather, it
deforms the squeezing kernel. If the pure-GR kernel is denoted by
\(\mathcal G^{(0)}_{ij}\), the Chern--Simons contribution gives
\begin{equation}
\mathcal G_{ij}^{\rm GR+CS}
=
\mathcal G_{ij}^{(0)}
+
\delta \mathcal G_{ij}^{\rm CS}.
\end{equation}
The Takagi decomposition of the deformed kernel is
\begin{equation}
\mathcal G^{\rm GR+CS}
=
\widetilde U\,
{\rm diag}
\left(
{\cal R}_A e^{i\Phi_A}
\right)
\widetilde U^T ,
\label{eq:TakagiCS2}
\end{equation}
where \(A\) labels the Takagi supermodes of the deformed kernel,
\({\cal R}_A\ge0\) are the corresponding squeezing amplitudes, and
\(\Phi_A\) are their squeezing phases.
The \(\omega\)-parameter is therefore obtained from the same open-system
formula as before, but with the modified Takagi data.
\vskip .2cm
\paragraph{Near-resonant behaviour and Markovian limit:}

Near resonance, $|\delta_\alpha t|\ll 1$, the time kernel reduces to
$\mathcal{F}_\alpha(t) \to t^2/2$: each near-resonant supermode contributes a
quadratic-in-time growth of $\omega$, the direct analogue of Zeno-regime
coherent accumulation.

In the Markovian limit one extends the memory integral to infinity using
$\int_0^\infty d\tau\,e^{-i\delta_\alpha\tau}
= \pi\delta(\delta_\alpha) - i\mathcal{P}\delta_\alpha^{-1}$,
separating the reduced dynamics into a dissipative resonant part
$\propto\pi\delta(\Omega_\alpha-\Delta E)$ and a dispersive Lamb-shift
correction $\propto\mathcal{P}(\Omega_\alpha-\Delta E)^{-1}$:
\begin{equation}
\label{MarkovApprox}
\omega_{\rm Markov}(t)
\sim
t\sum_\alpha g_\alpha^{(1)}g_\alpha^{(2)}\,M_\alpha\,\mathcal{C}_{AS}
\left[\pi\delta(\Omega_\alpha-\Delta E)
- i\,\mathcal{P}\frac{1}{\Omega_\alpha-\Delta E}\right].
\end{equation}
The transition from quadratic to linear-in-$t$ accumulation occurs on the
timescale set by the bath correlation time, i.e.\ the inverse bandwidth of the
gravitational radiation spectrum.

The dominant contribution to $\omega$ comes from supermodes satisfying
\begin{equation}
\Omega_\alpha \simeq \Delta E,
\end{equation}
the direct analogue of the phase-matching condition in nonlinear optics. In the
Kerr plus axion-cloud scenario, the graviton supermode frequencies are set by
axion annihilation ($\Omega\sim 2\omega_a$), so the strength of the effect is
controlled by how closely one of these frequencies approaches the meson
transition scale $\Delta E$.

During the derivation of the \(\omega\)-effect it is important to distinguish carefully between three different levels of description: \begin{enumerate} \item the underlying field-theoretic energy--momentum tensor \(T^{\mu\nu}(x)\); \item the graviton-mode-projected operator \({\cal T}_\alpha\); \item the effective two-level operator \(X_\alpha\) acting on the \(\{|K_S\rangle,|K_L\rangle\}\) sector. \end{enumerate} Confusing these three objects obscures the origin of the transition matrix elements responsible for the \(\omega\)-effect. 

\paragraph{From the stress tensor to graviton-mode operators:} The fundamental interaction is given in \eqref{eq:Sint_Tmunu}, \eqref{Hint1}, \eqref{Hint2}. After expanding the graviton field in Takagi supermodes, the operator \({\cal T}_\alpha\) still acts on the full kaon Hilbert space and contains all information about the graviton wavefunction, the meson wavefunctions, and their spatial overlap. 

The physically relevant quantities are the one-particle matrix elements   \begin{equation} (X_\alpha)_{ij}(\mathbf p',\mathbf p) = \frac12 \int d^3x\, {\cal U}^{(\alpha)}_{\mu\nu}(x) \langle K_i(\mathbf p')| T^{\mu\nu}(x) |K_j(\mathbf p)\rangle  \label{eq:Xij_explicit},\qquad i,j=S,L.  \end{equation} At this stage the momentum dependence remains explicit.  For the discussion of the \(\omega\)-effect one restricts attention to the neutral-kaon propagation subspace spanned by \(\{|K_S\rangle,|K_L\rangle\}\). Suppressing the momentum labels, the projected operator may be written as \begin{equation} X_\alpha = \begin{pmatrix} (X_\alpha)_{SS} & (X_\alpha)_{SL}\\[2mm] (X_\alpha)_{LS} & (X_\alpha)_{LL} \end{pmatrix}. \end{equation} Equivalently, \begin{equation} X_\alpha = c_0\mathbf 1 + c_1\sigma_1 + c_2\sigma_2 + c_3\sigma_3, \end{equation} or \begin{equation} X_\alpha = c_0\mathbf 1 + c_3\sigma_3 + c_+\sigma_+ + c_-\sigma_-, \end{equation} with \begin{equation} c_+ = c_1-i c_2, \qquad c_- = c_1+i c_2. \end{equation} The coefficients \(c_\pm\) are the off-diagonal matrix elements, \begin{equation} c_+ = (X_\alpha)_{LS}, \qquad c_- = (X_\alpha)_{SL}, \end{equation} and therefore measure transitions between \(K_S\) and \(K_L\). 

For a purely free diagonal theory one has \begin{equation} \langle K_S(\mathbf p')| T^{\mu\nu}_{\rm free}(x) |K_L(\mathbf p)\rangle = 0, \end{equation} implying \begin{equation} c_+=c_-=0. \end{equation} The physical neutral-kaon system is not free.  The non-zero transition coefficients \(c_\pm\) therefore originate from the weak-interaction flavour structure of the neutral-kaon system. The squeezed graviton bath does not generate this flavour mixing; rather, it acts on a system whose propagation eigenstates have already been determined by weak interactions.

The quantity entering the \(\omega\)-effect is naturally \begin{equation} {\cal C}_{AS}^{(\alpha)} = \langle S| {\cal T}_\alpha^{(1)} {\cal T}_\alpha^{(2)} |A\rangle . \label{eq:CAS_exact} \end{equation} Equation~(\ref{eq:CAS_exact}) contains the full dependence on graviton and kaon wavefunctions, momentum transfer, detector geometry and flavour structure.The superscripts \((1)\) and \((2)\) label the two mesons.
On the two-kaon Hilbert space
\(
{\cal H}_K^{(1)}\otimes{\cal H}_K^{(2)}
\),
the projected graviton-coupling operator acts as
\begin{equation}
{\cal T}_\alpha^{(1)}
=
{\cal T}_\alpha\otimes{\bf 1},
\qquad
{\cal T}_\alpha^{(2)}
=
{\bf 1}\otimes{\cal T}_\alpha.
\end{equation}
The quantity
\begin{equation}
{\cal C}_{AS}^{(\alpha)}
=
\langle S|
{\cal T}_\alpha^{(1)}
{\cal T}_\alpha^{(2)}
|A\rangle
\end{equation}
is therefore the matrix element connecting the antisymmetric and
symmetric two-meson sectors through the simultaneous action of the
graviton mode \(\alpha\) on both mesons.

The exact expression for the \(\omega\)-parameter is therefore \begin{equation} \omega(t) = - \sum_\alpha M_\alpha {\cal C}_{AS}^{(\alpha)} F_\alpha(t), \end{equation} where \begin{equation} M_\alpha = - e^{i\phi_\alpha} \sinh r_\alpha\cosh r_\alpha . \end{equation} Only if one makes the additional approximation \begin{equation} {\cal T}_\alpha = g_\alpha X, \end{equation} does the matrix element factorise: \begin{equation} {\cal C}_{AS}^{(\alpha)} = g_\alpha^{(1)} g_\alpha^{(2)} \langle S| X^{(1)}X^{(2)} |A\rangle . \end{equation} Defining \begin{equation} {\cal C}_{AS} = \langle S| X^{(1)}X^{(2)} |A\rangle , \end{equation} one obtains the factorised form \begin{equation} \omega(t) = - \sum_\alpha g_\alpha^{(1)} g_\alpha^{(2)} M_\alpha {\cal C}_{AS} F_\alpha(t). \end{equation} It is important to emphasise that this factorisation is an additional approximation. In the exact treatment all wavefunction overlap information is contained in \({\cal C}_{AS}^{(\alpha)}\). \paragraph{Physical interpretation:} The \(\omega\)-effect requires two independent ingredients: \begin{enumerate} \item weak-interaction-induced flavour mixing, encoded in the transition coefficients \(c_\pm\); \item anomalous squeezed-graviton correlators \(\langle\mathbb b_\alpha\mathbb b_\alpha\rangle\). \end{enumerate} Neither ingredient alone is sufficient. A weakly mixed kaon system in a thermal gravitational bath gives no \(\omega\)-effect because the anomalous correlators vanish. Conversely, a squeezed graviton bath coupled to a purely diagonal kaon operator (\(c_\pm=0\)) cannot generate the required transitions between the antisymmetric and symmetric two-meson sectors. The \(\omega\)-effect therefore arises from the interplay between weak flavour mixing and graviton squeezing.


We comment further on this aspect of our analysis. 
The explicit expression for $\omega(t)$ derived above was obtained within the effective
two-level description of the neutral-kaon system. In this approximation
the matrix element governing the transition between the antisymmetric and
exchange-symmetric sectors was reduced to a simple flavour factor
multiplied by a phase determined by the \(K_S\)--\(K_L\) energy
splitting.

More generally, however, the quantity entering the Born kernel is
\begin{equation}
{\cal C}_{AS}^{(\alpha)}(t_1,t_2)
=
\left\langle S\left|
{\cal T}_{\alpha}^{(1)}(t_1)
{\cal T}_{\alpha}^{(2)}(t_2)
\right|A\right\rangle ,
\end{equation}
where \({\cal T}_\alpha\) denotes the graviton-mode-projected meson
operator. This object contains the full dependence on the meson
wave packets, momentum transfer, weak-interaction dynamics, and graviton
mode structure.
The simplified expression used in the main text assumed
\begin{equation}
C^{(K)}_{AS}(t_1,t_2)
\simeq
C^{(K)}_{AS}
e^{+i\Delta\lambda(t_1+t_2)},
\end{equation}
with
\begin{equation}
\Delta\lambda
=
\lambda_L-\lambda_S
=
\Delta m_K-\frac{i}{2}\Delta\Gamma_K ,
\end{equation}
which should  be
viewed as the leading approximation to the more general response
function \({\cal C}_{AS}^{(\alpha)}(t_1,t_2)\). The central physical
conclusion remains unchanged: the \(\omega\)-effect is controlled by the
product of the anomalous graviton correlator and the matrix element
connecting the antisymmetric and symmetric meson sectors.

\subsection{Markov approximation}

The Markov approximation used in \eqref{MarkovApprox} more generally leads to a Lindbladian form of the master equation \cite{Lindblad:1976,Gorini:1975nb}.
Writing $\tau=t-s$, \eqref{SecondOrderMaster} becomes
\begin{equation}
\dot\rho_M(t)
=
-\int_0^t d\tau\,
{\rm Tr}_{\rm grav}
\left[
H_I(t),
\left[
H_I(t-\tau),
\rho_M(t-\tau)\otimes\rho_{\rm grav}
\right]
\right].
\end{equation}
The Markov approximation assumes that the graviton bath correlation time
is much shorter than the timescale on which the meson density matrix
changes. Hence
\begin{equation}
\rho_M(t-\tau)
\simeq
\rho_M(t),
\end{equation}
and one extends the upper limit of the memory integral:
\begin{equation}
\int_0^t d\tau
\longrightarrow
\int_0^\infty d\tau .
\end{equation}
This gives the Born--Markov master equation
\begin{equation}
\dot\rho_M(t)
=
-\int_0^\infty d\tau\,
{\rm Tr}_{\rm grav}
\left[
H_I(t),
\left[
H_I(t-\tau),
\rho_M(t)\otimes\rho_{\rm grav}
\right]
\right].
\end{equation}
Now write the interaction Hamiltonian as
\begin{equation}
H_I(t)
=
\sum_\alpha
\left[
S_\alpha(t)B_\alpha(t)
+
S_\alpha^\dagger(t)B_\alpha^\dagger(t)
\right],
\end{equation}
where $S_\alpha$ acts on the two-meson Hilbert space and $B_\alpha$ acts
on the graviton bath. In the Takagi basis one may identify
\begin{equation}
B_\alpha(t)=\mathbb{b}_\alpha e^{-i\Omega_\alpha t}.
\end{equation}
The bath trace produces correlation functions such as
\begin{equation}
C_{\alpha\beta}^{(+)}(\tau)
=
\langle
B_\alpha(t)B_\beta^\dagger(t-\tau)
\rangle,
\end{equation}
\begin{equation}
C_{\alpha\beta}^{(-)}(\tau)
=
\langle
B_\alpha^\dagger(t)B_\beta(t-\tau)
\rangle,
\end{equation}
and, for a squeezed bath, anomalous correlators
\begin{equation}
M_{\alpha\beta}(\tau)
=
\langle
B_\alpha(t)B_\beta(t-\tau)
\rangle ,
\end{equation}
\begin{equation}
M_{\alpha\beta}^*(\tau)
=
\langle
B_\alpha^\dagger(t)B_\beta^\dagger(t-\tau)
\rangle .
\end{equation}
For a Takagi-diagonal squeezed vacuum,

\begin{equation}
M_{\alpha\beta}(\tau)
=
-\delta_{\alpha\beta}
e^{i\phi_\alpha}
\sinh r_\alpha\cosh r_\alpha
e^{-i\Omega_\alpha(2t-\tau)} .
\end{equation}
To obtain a time-independent Lindblad generator~\cite{Lindblad:1976,Gorini:1975nb}, one decomposes the
system operator into eigenoperators of the meson Hamiltonian:
\begin{equation}
S_\alpha(t)
=
\sum_\omega
S_\alpha(\omega)e^{-i\omega t},
\end{equation}
where
\begin{equation}
[H_M,S_\alpha(\omega)]
=
-\omega S_\alpha(\omega).
\end{equation}
The  rotating-wave, approximation keeps only terms with
matching Bohr frequencies.The resulting master equation takes the
Lindblad form~\cite{Lindblad:1976,Gorini:1975nb}
\begin{equation}
\dot\rho_M
=
-i[H_M+H_{\rm LS},\rho_M]
+
\mathcal D_{\rm N}[\rho_M]
+
\mathcal D_{\rm sq}[\rho_M].
\end{equation}
The normal dissipator is
\begin{align}
\mathcal D_{\rm N}[\rho_M]
=
\sum_{\alpha,\omega}
\gamma_\alpha(\omega)
\left[
S_\alpha(\omega)\rho_M S_\alpha^\dagger(\omega)
-
\frac12
\left\{
S_\alpha^\dagger(\omega)S_\alpha(\omega),
\rho_M
\right\}
\right]
\nonumber\\
+
\sum_{\alpha,\omega}
\gamma_\alpha(-\omega)
\left[
S_\alpha^\dagger(\omega)\rho_M S_\alpha(\omega)
-
\frac12
\left\{
S_\alpha(\omega)S_\alpha^\dagger(\omega),
\rho_M
\right\}
\right].
\end{align}
The squeezed, or anomalous, dissipator contains terms of the form
\begin{align}
\mathcal D_{\rm sq}[\rho_M]
=
-\sum_{\alpha,\omega}
m_\alpha(\omega)
\left[
S_\alpha(\omega)\rho_M S_\alpha(-\omega)
-
\frac12
\left\{
S_\alpha(-\omega)S_\alpha(\omega),
\rho_M
\right\}
\right]
\nonumber\\
-\sum_{\alpha,\omega}
m_\alpha^*(\omega)
\left[
S_\alpha^\dagger(\omega)\rho_M S_\alpha^\dagger(-\omega)
-
\frac12
\left\{
S_\alpha^\dagger(-\omega)S_\alpha^\dagger(\omega),
\rho_M
\right\}
\right].
\end{align}

The rates are Fourier transforms of the bath correlators:
\begin{equation}
\gamma_\alpha(\omega)
=
\int_{-\infty}^{+\infty}d\tau\,
e^{i\omega\tau}
C_\alpha^{(+)}(\tau),
\end{equation}
and the squeezed coefficients are
\begin{equation}
m_\alpha(\omega)
=
\int_{-\infty}^{+\infty}d\tau\,
e^{i\omega\tau}
M_\alpha(\tau).
\end{equation}
Their imaginary principal-value parts contribute to the Lamb-shift
Hamiltonian,
\begin{equation}
H_{\rm LS}
=
\sum_{\alpha,\omega}
\Delta_\alpha(\omega)
S_\alpha^\dagger(\omega)S_\alpha(\omega)
+
\cdots .
\end{equation}

For the neutral-meson problem, the distinctive feature of the squeezed
graviton bath is the presence of anomalous correlators
\(
\langle \mathbb b_\alpha \mathbb b_\beta\rangle
\).
These give rise to additional terms in the dissipator that are absent for
vacuum, thermal, or coherent baths.
The anomalous part of the dissipator contains products of
two meson transition operators,
\begin{equation}
X^{(1)}X^{(2)},
\end{equation}
weighted by the anomalous graviton correlators. The precise operator
structure depends on the form of the projected meson operator \(X\), but
the important point is that these terms generate matrix elements between
the exchange-antisymmetric and exchange-symmetric sectors of the
two-meson Hilbert space.

The reduced density matrix is initially taken to be
\begin{equation}
\rho_M(0)
=
|A\rangle\langle A|,
\end{equation}
where
\begin{equation}
|A\rangle
=
\frac1{\sqrt2}
\left(
|K_SK_L\rangle
-
|K_LK_S\rangle
\right).
\end{equation}
The quantity of interest is the off-diagonal density-matrix element
\begin{equation}
\omega(t)
=
\langle S|
\rho_M(t)
|A\rangle ,
\end{equation}
with
\begin{equation}
|S\rangle
=
\frac1{\sqrt2}
\left(
|K_SK_L\rangle
+
|K_LK_S\rangle
\right).
\end{equation}

For short times,
\begin{equation}
\rho_M(t)
=
\rho_M(0)
+
t\,\mathcal L[\rho_M(0)]
+
\mathcal O(t^2),
\end{equation}
so that
\begin{equation}
\omega(t)
=
t\,
\langle S|
\mathcal L_{\rm sq}
\!\left[
|A\rangle\langle A|
\right]
|A\rangle
+
\mathcal O(t^2).
\end{equation}
Thus the \(\omega\)-effect is generated whenever the squeezed-bath part
of the Lindblad generator has non-vanishing matrix elements between the
antisymmetric and symmetric exchange sectors.

\section{Application to the Neutral Kaon System}
\label{sec:kaon}

\subsection{Kaon mass and width parameters}

The neutral kaon system was introduced in Section~\ref{sec:interaction}
as the primary application of our general framework. The propagation
eigenstates $|K_S\rangle$ and $|K_L\rangle$ are defined there in
terms of the flavour states $|K^0\rangle$ and $|\bar K^0\rangle$. For
the kaon system the relevant energy gap is complex,
\begin{equation}
\Delta\lambda = \lambda_L - \lambda_S
= \Delta m_K - \tfrac{i}{2}\Delta\Gamma_K,
\end{equation}
where $\lambda_{S,L} = m_{S,L} - \frac{i}{2}\Gamma_{S,L}$ are the
complex energies, $\Delta m_K = m_L - m_S$ is the mass difference, and
$\Delta\Gamma_K = \Gamma_L - \Gamma_S$ is the width difference. This
complex splitting replaces $\Delta E$ in the general
formula~\eqref{eq:omega_main}.

\subsection{Modified initial state and time evolution}

In the absence of the gravitational environment the initial kaon state from
$\phi\to K^0\bar{K}^0$ is
\begin{equation}
|i\rangle
=
\frac{1}{\sqrt2}
\bigl(|K_S\rangle|K_L\rangle - |K_L\rangle|K_S\rangle\bigr).
\end{equation}
The squeezed graviton bath modifies this to
\begin{equation}
|i\rangle_\omega
=
\mathcal N
\Bigl[
|K_S\rangle|K_L\rangle - |K_L\rangle|K_S\rangle
+\omega\bigl(|K_S\rangle|K_S\rangle - |K_L\rangle|K_L\rangle\bigr)
\Bigr],
\end{equation}
and the two-kaon state at times $(t_1,t_2)$ is
\begin{align}
|\psi(t_1,t_2)\rangle
&=
\frac{1}{\sqrt2}
\Bigl(
e^{-i\lambda_S t_1}e^{-i\lambda_L t_2}|K_S\rangle|K_L\rangle
-e^{-i\lambda_L t_1}e^{-i\lambda_S t_2}|K_L\rangle|K_S\rangle
\Bigr)
\nonumber\\
&\quad
+\omega\Bigl(
e^{-i\lambda_S(t_1+t_2)}|K_S\rangle|K_S\rangle
-e^{-i\lambda_L(t_1+t_2)}|K_L\rangle|K_L\rangle
\Bigr).
\end{align}

\subsection{Decay intensity, experimental observable, and kaon-specific $\omega$}
In a \(\phi\)-factory the entangled neutral-kaon pair is observed through
the decay products of the two kaons. We denote by \(f_1\) the final state
into which the first kaon decays, and by \(f_2\) the final state into
which the second kaon decays. The corresponding decay times are \(t_1\)
and \(t_2\), and the experimentally measured distribution is usually
written in terms of \(\Delta t=t_1-t_2\).

The case \(f_1=f_2=\pi^+\pi^-\) means that both kaons decay into the same
charged two-pion final state,
\[
K\to \pi^+\pi^- .
\]
Here \(\pi^+\) and \(\pi^-\) denote the positively and negatively charged
pions. This channel is especially sensitive to the \(\omega\)-effect
because, in the absence of an exchange-symmetric admixture, the
antisymmetric EPR state suppresses identical decay configurations at
\(\Delta t=0\).For identical final states \(f_1=f_2=f\), the decay amplitude arising
from the antisymmetric state is
\begin{equation}
{\cal A}_A(t_1,t_2)
\propto
e^{-i\lambda_S t_1}
e^{-i\lambda_L t_2}
-
e^{-i\lambda_L t_1}
e^{-i\lambda_S t_2},
\label{eq:AA_decay}
\end{equation}
where
\begin{equation}
\lambda_{S,L}
=
m_{S,L}
-\frac{i}{2}\Gamma_{S,L}.
\end{equation}
The corresponding intensity is
\begin{equation}
I_A(\Delta t)
\propto
|{\cal A}_A|^2 .
\end{equation}
A straightforward calculation yields
\begin{align}
I_A(\Delta t)
&\propto
e^{-\Gamma_L|\Delta t|}
+
e^{-\Gamma_S|\Delta t|}
\nonumber\\
&\quad
-
2
e^{-(\Gamma_S+\Gamma_L)|\Delta t|/2}
\cos(\Delta m\,\Delta t),
\label{eq:IA}
\end{align}
where
\begin{equation}
\Delta m
=
m_L-m_S .
\end{equation}

The total decay amplitude becomes
\begin{equation}
{\cal A}
=
{\cal A}_A
+
\omega {\cal A}_S ,
\end{equation}
where
\begin{equation}
{\cal A}_S(t_1,t_2)
\propto
e^{-i\lambda_S t_1}
e^{-i\lambda_L t_2}
+
e^{-i\lambda_L t_1}
e^{-i\lambda_S t_2}.
\end{equation}

The observable intensity is therefore
\begin{align}
I(\Delta t)
&\propto
|{\cal A}|^2
\nonumber\\
&=
|{\cal A}_A|^2
+
|\omega|^2
|{\cal A}_S|^2
+
2\,\mathrm{Re}
\!\left(
\omega\,
{\cal A}_S
{\cal A}_A^*
\right).
\end{align}

Substituting Eq.~\eqref{eq:IA} gives
\begin{align}
I(\Delta t)
&\propto
e^{-\Gamma_L|\Delta t|}
+
e^{-\Gamma_S|\Delta t|}
-
2e^{-(\Gamma_S+\Gamma_L)|\Delta t|/2}
\cos(\Delta m\,\Delta t)
\nonumber\\
&\quad
+
|\omega|^2 e^{-\Gamma_S|\Delta t|}
+
2\,\mathrm{Re}(\omega)\,
F(\Delta t),
\label{eq:Iomega}
\end{align}
where \(F(\Delta t)\) denotes the interference term
\begin{equation}
F(\Delta t)
\propto
{\rm Re}
\!\left(
{\cal A}_S
{\cal A}_A^*
\right).
\end{equation}

The physical significance of the \(\omega\)-effect becomes apparent near
\(\Delta t=0\). From Eq.~\eqref{eq:AA_decay},
\begin{equation}
{\cal A}_A(t,t)=0,
\end{equation}
so that the antisymmetric EPR state suppresses identical simultaneous
decays. A non-zero \(\omega\) introduces a symmetric component,
destroying this exact cancellation and thereby modifying the decay
distribution in the region \(\Delta t\approx0\). This is why the
\(\pi^+\pi^-\), \(\pi^+\pi^-\) channel provides one of the most
sensitive probes of the \(\omega\)-effect~\cite{Bernabeu:2004prl}.
The sensitivity of the \(\pi^+\pi^-,\pi^+\pi^-\) channel is enhanced
because the standard \(K_L\to\pi^+\pi^-\) amplitude is suppressed by the
small CP-violating parameter~\cite{Amelino-Camelia:2010cem}
\[
\eta_{+-}
=
\frac{A(K_L\to\pi^+\pi^-)}
     {A(K_S\to\pi^+\pi^-)}
\simeq 2.2\times10^{-3}.
\]
The \(\omega\)-induced symmetric component can contribute through
\(K_S K_S\)-like terms, which are not suppressed by this small factor.
Thus the relative size of the observable interference is effectively
controlled by \(|\omega|/|\eta_{+-}|\).

Including the finite lifetimes of the neutral-kaon propagation eigenstates,
the real energy splitting \(\Delta E\) appearing in
Eq.~\eqref{eq:omega_main} should be replaced by the complex difference
\begin{equation}
\Delta\lambda
=
\lambda_L-\lambda_S
=
\Delta m_K-\frac{i}{2}\Delta\Gamma_K,
\qquad
\lambda_{S,L}
=
m_{S,L}-\frac{i}{2}\Gamma_{S,L}.
\end{equation}
The kaon expression therefore becomes
\begin{equation}
\omega_K(t)
=
-\frac12
\sum_\alpha
g_\alpha^{(1)}g_\alpha^{(2)}
e^{i\phi_\alpha}
\sinh r_\alpha\cosh r_\alpha\,
\mathcal{C}_{AS}^{(K)}
\frac{\bigl[1-e^{-i(\Omega_\alpha-\Delta\lambda)t}\bigr]^2}
{(\Omega_\alpha-\Delta\lambda)^2}.
\end{equation}
Here \(\mathcal{C}_{AS}^{(K)}\) denotes the kaon-sector matrix element
defined in Appendix~\ref{app:Tmunu}.

The resonance condition is correspondingly controlled by the complex
detuning
\begin{equation}
\delta_\alpha
=
\Omega_\alpha-\Delta\lambda .
\end{equation}
Equivalently, appreciable enhancement occurs when
\begin{equation}
|\Omega_\alpha-\Delta m_K|
\lesssim
\frac12|\Delta\Gamma_K|.
\end{equation}
Thus the relevant spectral comparison is between the Takagi supermode
frequency \(\Omega_\alpha\) and the complex kaon splitting
\(\Delta\lambda\), rather than the naive comparison
\(2\mu_a\simeq\Delta m_K\). Off-resonant modes still contribute
dispersively, with their effect controlled by the squeezing amplitude,
the mode overlap, and the finite environmental correlation time. This resonance factor should be distinguished from the enhancement
appearing in stochastic D-foam estimates \cite{Bernabeu:2006} of the \(\omega\)-effect, where
the relevant scale is the ratio of a flavour-changing recoil fluctuation
to the neutral-meson level splitting.

\subsection{Estimates}
\label{sec:estimate}
The expression \eqref{eq:omega_main} provides the formal prediction for
the \(\omega\)-parameter. It is nevertheless useful to identify the
physical ingredients controlling its magnitude. For a dominant
near-resonant Takagi supermode \(\alpha=\ast\), the result may be read as
\begin{equation}
|\omega|
\sim
\sqrt{N_\ast(N_\ast+1)}
\,|\Xi_\ast|,
\end{equation}
where
\begin{equation}
N_\ast=\sinh^2 r_\ast
\end{equation}
is the occupation number of the squeezed graviton supermode and
\(\Xi_\ast\) denotes the combined effect of the graviton--meson coupling,
the resonance denominator, the finite interaction time, and the
antisymmetric--symmetric transition matrix element.

The crucial enhancement originates from the anomalous correlator of the
squeezed state,
\begin{equation}
\langle
\mathbb b_\ast \mathbb b_\ast
\rangle
=
-
e^{i\phi_\ast}
\sinh r_\ast\cosh r_\ast ,
\end{equation}
whose magnitude is
\begin{equation}
\left|
\langle
\mathbb b_\ast \mathbb b_\ast
\rangle
\right|
=
\sqrt{N_\ast(N_\ast+1)}.
\end{equation}
Thus
\begin{equation}
\sqrt{N_\ast(N_\ast+1)}
\sim
\begin{cases}
\sqrt{N_\ast},
&
N_\ast\ll1,
\\[2mm]
N_\ast,
&
N_\ast\gg1.
\end{cases}
\end{equation}
This contribution vanishes identically in the absence of squeezing and
has no analogue in a thermal bath.\footnote{This confirms the conclusion
of \cite{Bernabeu:2006}, which argued that thermal-bath-like
environments do not generate an \(\omega\)-effect.}

It is useful to separate the universal Planck suppression from the
remaining dynamical factors. Since the graviton couples through
\begin{equation}
{\cal L}_{\rm int}
\sim
\frac{1}{M_{\rm Pl}}
h_{\mu\nu}T^{\mu\nu},
\end{equation}
the corresponding dimensionless coupling is controlled by
\begin{equation}
\frac{E_K}{M_{\rm Pl}},
\end{equation}
where \(E_K\) is the characteristic kaon energy. One may therefore write
the schematic estimate
\begin{equation}
|\omega|
\sim
\Xi\,
\sqrt{N_{\rm eff}}
\left(
\frac{E_K}{M_{\rm Pl}}
\right)^2 ,
\label{eq:omega_parametric}
\end{equation}
where \(N_{\rm eff}\) is an effective occupation number of the
near-resonant graviton supermodes,
\begin{equation}
N_{\rm eff}
=
\sum_{\alpha\in{\rm res}}
{\cal O}_\alpha^2 N_\alpha ,
\end{equation}
and \(\Xi\) absorbs the resonance enhancement, the finite interaction
time, the graviton-mode overlap, and the matrix element
\({\cal C}_{AS}^{(\alpha)}\). Here \({\cal O}_\alpha\) denotes the overlap of the kaon transition
kernel with the graviton supermode \(\alpha\). Modes that couple weakly
to the antisymmetric--symmetric transition have \({\cal O}_\alpha\ll1\),
while strongly coupled modes have \({\cal O}_\alpha={\cal O}(1)\).

For neutral kaons,
\begin{equation}
E_K\sim0.5~{\rm GeV},
\qquad
M_{\rm Pl}\sim1.2\times10^{19}~{\rm GeV},
\end{equation}
so that
\begin{equation}
\left(
\frac{E_K}{M_{\rm Pl}}
\right)^2
\sim10^{-39}.
\end{equation}
This estimate should not be interpreted as the final prediction of the
model. It merely quantifies the weakness of the microscopic
gravitational interaction. The actual value of \(\omega\) depends on the
resonant structure of the graviton spectrum, the squeezing amplitudes,
the coherence properties of the environment, and the overlap between the
graviton supermodes and the neutral-meson sector.

A phenomenologically relevant effect therefore requires some combination
of:
\begin{enumerate}
\item resonant matching,
      \(\Omega_\ast\simeq \Delta\lambda\);
\item large graviton occupation number \(N_{\rm eff}\);
\item strong squeezing;
\item favourable overlap between the graviton spectrum and the
      antisymmetric--symmetric transition kernel.
\end{enumerate}

\paragraph{Astrophysical propagation effects:}
The mechanism described in this work is formally independent of the
location of the squeezed graviton source. What enters the reduced meson
dynamics is the graviton correlation function evaluated at the detector.
In principle, sizeable values of \(\omega\) could arise if the entangled
kaons were produced in the immediate vicinity of a rotating black hole
surrounded by a highly occupied superradiant axion cloud, where
\(N_{\rm eff}\gg1\). Such environments are not directly accessible to
experiment.

A more realistic possibility is that a terrestrial meson factory is
irradiated by gravitational waves containing a squeezed graviton
component emitted by a distant astrophysical source. In this case the
gravitational environment is diluted by propagation over astronomical
distances. A useful characterisation of the incoming radiation is the
gravitational-wave energy flux,
\begin{equation}
F_{\rm GW}
\sim
\frac{\pi c^3}{4G}
f^2 h_0^2,
\end{equation}
where \(h_0\) is the strain amplitude at Earth and \(f\) is the wave
frequency. For representative continuous-wave values
\begin{equation}
h_0\sim10^{-25},
\qquad
f\sim100~{\rm Hz},
\end{equation}
one finds
\begin{equation}
F_{\rm GW}
\sim
3\times10^{-11}
\left(
\frac{h_0}{10^{-25}}
\right)^2
\left(
\frac{f}{100~{\rm Hz}}
\right)^2
{\rm W\,m^{-2}}.
\end{equation}

The corresponding graviton number flux is
\begin{equation}
\Phi_g
=
\frac{F_{\rm GW}}{h_{\rm P}f}
\sim
5\times10^{20}
\left(
\frac{h_0}{10^{-25}}
\right)^2
\left(
\frac{f}{100~{\rm Hz}}
\right)
{\rm m^{-2}s^{-1}}.
\end{equation}

Although this number flux may appear large, each graviton carries an
extremely small energy and the coupling to neutral mesons remains
Planck-suppressed. Consequently, an astrophysical squeezed graviton
background is not expected to act as a strong local environment for a
terrestrial \(\phi\)-factory unless additional enhancement mechanisms
are present, most importantly resonance between the graviton supermode
frequencies and the kaon transition scale \(\Delta\lambda\).

Whether these enhancement conditions can be simultaneously realised for
a realistic astrophysical source remains an open question requiring a
full Kerr-background calculation of the graviton spectrum and its
coupling to the neutral-meson sector.


\section{Conclusions}
\label{sec:conclusions}

We have presented the first microscopic derivation of the $\omega$-effect in
neutral meson systems, arising from their interaction with a multimode squeezed
graviton background sourced by axion clouds around Kerr black holes. \cbl{In this environment parent particles  such as $\phi$-particles (for the case of neutral Kaons), are assumed to be produced by appropriate cosmic processes}.

The key structural ingredient is the squeezing kernel $\mathcal{G}_{ij}$, which encodes
correlated graviton pair production by the axion cloud. Its Takagi decomposition
diagonalises the multimode state into independent supermodes, each contributing
additively to the induced meson decoherence. The $\omega$-parameter emerges as a
coherent sum over supermode contributions [Eq.~\eqref{eq:omega_main}], weighted
by their squeezing amplitudes and phases, with the dominant contribution from
near-resonant modes. In the non-relativistic limit the kernel is peaked on
anti-collinear graviton pairs, directly analogous to phase-matching in nonlinear
optics; in the relativistic Kerr geometry this constraint is relaxed.

The CPT violation is not a breakdown of fundamental CPT symmetry at the
Lagrangian level but an emergent consequence of open-system evolution, as
guaranteed by Wald's theorem: the microscopic meson--graviton Hamiltonian can
itself be CPT invariant. The Markovian time-kernel analysis clarifies how
resonant matching between graviton pair energies and the effective meson
transition frequency $\Delta\lambda = \Delta m_K-\frac{i}{2}\Delta\Gamma_K$
governs the magnitude of the effect.

The precise quantum replacement for the stochastic D-foam variance is the
anomalous correlator of squeezed graviton supermodes
[Eq.~\eqref{eq:dictionary}], which vanishes identically for any thermal or
classical gravitational noise. Squeezing is therefore both necessary and
sufficient for the $\omega$-effect in this framework. As discussed in
Section~\ref{sec:estimate}, a phenomenologically relevant value of $|\omega|$
requires resonant matching of a graviton supermode with the kaon transition
scale, large Takagi occupation numbers, and favourable spectral overlap; the
astrophysical dilution of the flux from a black-hole axion cloud poses a
substantial challenge that full Kerr-background calculations would need to
address.

If microscopic black holes were produced in collider environments, their Hawking
radiation would also admit a squeezed-state interpretation; however, the
characteristic frequencies would be set by the Hawking temperature and would
generically be far above the neutral-kaon transition scale, making the resonance
condition difficult to satisfy in the present $\omega$-effect framework.

\cbl{The squeezing-induced $\omega$-effect is an indirect way to detect squeezed quantum states of gravitons in gravitational waves.
Indeed, in the astrophysical set up of \cite{Dorlis:2025zzz,Dorlis:2025amf,Dorlis:2026gth} a direct detection of such states would be associated with the sub-Poissonian noise statistics in the detector (variance is smaller than the mean, $\langle N^2\rangle - \langle N \rangle^2 < \langle N \rangle$)~\cite{Parikh:2020nrd},\cite{Parikh:2020fhy}; in contrast  coherent graviton modes  exhibit super-Poissonian statistics. Classical GR does not produce sub-Poisson statistics, and thus a potential detection of sub-Possionian noise would imply the presence of quantum features in the gravitational wave. The induced $\omega$-effect on a neutral meson factory, struck by a gravitational wave containing squeezed graviton states, offers therefore complementary information (albeit suppressed) on the quantum nature of gravity.}

Several directions for future work are indicated. A full computation of
$\mathcal{G}_{ij}$ in the Kerr background~\cite{Dorlis:2025amf,Dorlis:2025zzz},
including relativistic corrections and greybody factors, is required for
quantitative predictions. Embedding the mechanism in a detector-level description
would connect it directly to observables at \cbl{future neutral-meson 
factories. Although here we chose explicitly the neutral-Kaon system for our study, the formalism extends naturally to other neutral-meson
systems (B mesons, D mesons) and to any quantum system sensitive to
environmental squeezing.}

\cbl{Last, but not least, the construction of a potential condensed-matter analogue  system of a rotating black-hole surrounded by long-lived (pseudo) scalar clouds, would be an important development, since such systems could lead to analogues for the emergence of an $\omega$-like effect in the Laboratory.
Finally, although the present work has focused on entangled neutral
mesons, the underlying mechanism is considerably more general. The key
ingredients are an entangled two-level system and an environment
possessing non-trivial pair correlations. In the neutral-kaon case this
manifests itself as an \(\omega\)-effect generated by a squeezed graviton
bath. However, closely related phenomena may arise in other quantum
systems. In particular, axion electrodynamics \cite{Eichhorn:2012qf},
\[
a\,F_{\mu\nu}\widetilde F^{\mu\nu},
\]
provides an electromagnetic analogue of the axion--Chern--Simons
gravitational coupling considered here with $a$ denoting the axion field.
The dual electromagnetic field-strength tensor is defined by
\begin{equation}
\widetilde F^{\mu\nu}
=
\frac12
\epsilon^{\mu\nu\rho\sigma}
F_{\rho\sigma},
\end{equation}
so that
\begin{equation}
F_{\mu\nu}\widetilde F^{\mu\nu}
=
\frac12
\epsilon^{\mu\nu\rho\sigma}
F_{\mu\nu}F_{\rho\sigma}
=
-4\,\mathbf E\!\cdot\!\mathbf B .
\end{equation}
The axion-electrodynamics interaction
\begin{equation}
\mathcal L_{\rm int}
=
\frac{g_{a\gamma\gamma}}{4}
a\,F_{\mu\nu}\widetilde F^{\mu\nu}
\end{equation}
is therefore a parity-odd coupling between the axion field and the
electromagnetic Pontryagin density, analogous to the gravitational
Chern--Simons interaction
\(b\,{}^\star RR\).\footnote{\cbl{There are important differences between the gauge-chiral-anomaly and the gravitational-anomaly terms, as we have discussed previously. Indeed, the former are topological, independent of the metric, while the latter exhibit a non-trivial variation with respect to the graviton field, yielding non-trivial contributions (Cotton tensor \eqref{eq:Cotton}) to the Einstein equations, implying energy exchange between the axion and the gravitational anomaly terms~\cite{jackiw,anomalies}.}}
Since axion backgrounds induce
helicity-dependent photon propagation and can generate non-classical
states of light, it is natural to ask whether entangled photon pairs may
exhibit an optical analogue of the \(\omega\)-effect, corresponding to a
small admixture of exchange-symmetric and exchange-antisymmetric Bell
states. Such systems would not probe CPT violation directly, but could
provide experimentally accessible laboratories in which the open-system
mechanism developed in this work may be investigated.
Such issues will be looked at in future works.}

\begin{acknowledgments}
 \cbl{The work of NEM and SS is supported in part by the Science and Technology Facilities Council (UK) under
grant ST/X000753/1.} 
N.E.M.\ also acknowledges participation in the COST Association Actions
CA21136 ``Addressing observational tensions in cosmology with systematics and
fundamental physics (CosmoVerse)'' and CA23130 ``Bridging high and low energies in
search of Quantum Gravity (BridgeQG)''.
\end{acknowledgments}

\appendix

\section{Calculation of \texorpdfstring{$\mathcal C_{AS}^{(K)}$}{CASK}
from the Neutral-Kaon Stress Tensor}
\label{app:Tmunu}

The meson-sector factor entering the $\omega$-effect is
\begin{equation}
\mathcal C_{AS}^{(K)}
= \langle S_\omega|X^{(1)}(t_1)X^{(2)}(t_2)|A\rangle,
\end{equation}
with $|A\rangle = \frac{1}{\sqrt2}(|K_SK_L\rangle-|K_LK_S\rangle)$ and
$|S_\omega\rangle = \frac{1}{\sqrt2}(|K_SK_S\rangle-|K_LK_L\rangle)$.

\subsection{Free stress tensor and its diagonal structure}
We show how the effective two-level operator \(X\)
used in the main text arises from the underlying field-theoretic
energy--momentum tensor.
We begin with the flavour doublet
\begin{equation}
\Phi(x)
=
\begin{pmatrix}
K^0(x)\\
\bar K^0(x)
\end{pmatrix},
\end{equation}
whose free dynamics is described by
\begin{equation}
\mathcal L
=
\partial_\mu\Phi^\dagger
\partial^\mu\Phi
-
\Phi^\dagger {\cal M}^2 \Phi .
\end{equation}

The associated Fock space is generated by creation operators
\begin{equation}
a_{K^0}^\dagger(\mathbf p),
\qquad
a_{\bar K^0}^\dagger(\mathbf p),
\end{equation}
acting on the vacuum:
\begin{equation}
|K^0(\mathbf p)\rangle
=
a_{K^0}^\dagger(\mathbf p)|0\rangle,
\qquad
|\bar K^0(\mathbf p)\rangle
=
a_{\bar K^0}^\dagger(\mathbf p)|0\rangle.
\end{equation}

The full Fock space contains arbitrary numbers of kaons,
\begin{equation}
{\cal F}
=
{\cal H}_0
\oplus
{\cal H}_1
\oplus
{\cal H}_2
\oplus
\cdots .
\end{equation}

In the present problem we are interested only in the \emph{one-particle}
neutral-kaon sector.

Weak interactions generate off-diagonal terms in the effective neutral-kaon
Hamiltonian,
\begin{equation}
H_K
=
M
-
\frac{i}{2}\Gamma ,
\end{equation}
through strangeness changing \(|\Delta S|=2\) processes.

Diagonalisation yields the propagation eigenstates
\begin{equation}
|K_S(\mathbf p)\rangle
=
p|K^0(\mathbf p)\rangle
+
q|\bar K^0(\mathbf p)\rangle ,
\end{equation}
\begin{equation}
|K_L(\mathbf p)\rangle
=
p|K^0(\mathbf p)\rangle
-
q|\bar K^0(\mathbf p)\rangle .
\end{equation}

These states span a two-dimensional flavour subspace at fixed momentum:
\begin{equation}
{\cal H}_{\mathbf p}
=
{\rm span}
\Bigl\{
|K_S(\mathbf p)\rangle,
|K_L(\mathbf p)\rangle
\Bigr\}.
\end{equation}

The graviton couples to the local energy--momentum tensor,
\begin{equation}
S_{\rm int}
=
\frac12
\int d^4x\,
h_{\mu\nu}(x)
T^{\mu\nu}(x).
\end{equation}
The operator \(T^{\mu\nu}(x)\) acts on the full Fock space.
To obtain an effective neutral-kaon description, we consider its
matrix elements between one-particle kaon states:
\begin{equation}
T^{\mu\nu}_{ij}(p',p)
=
\langle K_i(\mathbf p')|
T^{\mu\nu}(0)
|K_j(\mathbf p)\rangle,
\qquad
i,j=S,L.
\end{equation}

For a free diagonal theory one finds
\begin{equation}
T^{\mu\nu}_{ij}(p',p)
=
\delta_{ij}
\Bigl[
p'^\mu p^\nu
+
p'^\nu p^\mu
-
\eta^{\mu\nu}
(p'\!\cdot\!p-m_i^2)
\Bigr].
\end{equation}

Thus
\begin{equation}
\langle K_S(\mathbf p')|
T^{\mu\nu}(0)
|K_L(\mathbf p)\rangle
=
0
\end{equation}
in the free theory.

\subsection{Definition of the effective operator \(X\)}
The graviton field couples universally to the energy--momentum tensor of the meson system,
\begin{equation}\label{tmnmeson}
S_{\rm int}
=
\frac12
\int d^4x\,
h_{\mu\nu}(x)\,
T^{\mu\nu}(x).
\end{equation}
The graviton field is expanded in Takagi supermodes \eqref{TakagiExpansion}. The operator \({\cal T}_\alpha\) acts on the full kaon Hilbert space.
To obtain an effective description of neutral-kaon flavour dynamics, we
restrict attention to the one-particle sector spanned by the propagation
eigenstates
\begin{equation}
|K_S(\mathbf p)\rangle,
\qquad
|K_L(\mathbf p)\rangle.
\end{equation}
The projected operator is therefore defined through the matrix elements
\begin{equation}
(X_\alpha)_{ij}
=
\langle K_i(\mathbf p')|
{\cal T}_\alpha
|K_j(\mathbf p)\rangle,
\qquad
i,j=S,L.
\label{eq:Xproj}
\end{equation}
Explicitly,
\begin{equation}
(X_\alpha)_{ij}(\mathbf p',\mathbf p)
=
\frac12
\int d^3x\,
{\cal U}^{(\alpha)}_{\mu\nu}(x)
\langle K_i(\mathbf p')|
T^{\mu\nu}(x)
|K_j(\mathbf p)\rangle .
\label{eq:Xalpha_momentum}
\end{equation}

The diagonal entries describe graviton-induced phase shifts of the
\(K_S\) and \(K_L\) propagation eigenstates, whereas the off-diagonal
entries describe transitions between them. It is these off-diagonal
matrix elements that generate the flavour-changing coefficients entering
the effective two-level description.

To make this explicit, it is useful to separate the stress tensor into a
free part and a weak-interaction contribution,
\begin{equation}
T^{\mu\nu}
=
T^{\mu\nu}_{\rm free}
+
\delta T^{\mu\nu}_{\rm weak}.
\end{equation}
For the free theory one finds
\begin{equation}
\langle K_S(\mathbf p')|
T^{\mu\nu}_{\rm free}(0)
|K_L(\mathbf p)\rangle
=
0,
\end{equation}
so that
\begin{equation}
\langle K_S|
{\cal T}_{\alpha,{\rm free}}
|K_L\rangle
=
0.
\end{equation}
Hence a purely free gravitational coupling produces only diagonal
contributions.

The off-diagonal matrix elements originate from the weak-interaction
structure responsible for \(K^0\)-\(\bar K^0\) mixing:
\begin{equation}
\langle K_S|
{\cal T}_\alpha
|K_L\rangle
=
\frac12
\int d^3x\,
{\cal U}^{(\alpha)}_{\mu\nu}(x)\,
\langle K_S|
\delta T^{\mu\nu}_{\rm weak}(x)
|K_L\rangle .
\end{equation}
These matrix elements define the transition coefficients appearing in the
effective two-level operator.

Expanding \(X_\alpha\) in the Pauli basis,
\begin{equation}
X_\alpha
=
c_0\mathbf 1
+
c_1\sigma_1
+
c_2\sigma_2
+
c_3\sigma_3,
\end{equation}
or equivalently
\begin{equation}
X_\alpha
=
c_0\mathbf 1
+
c_3\sigma_3
+
c_+\sigma_+
+
c_-\sigma_-,
\end{equation}
with
\begin{equation}
c_+
=
c_1-i c_2,
\qquad
c_-
=
c_1+i c_2,
\end{equation}
one identifies
\begin{equation}
c_+
=
\langle K_L|X_\alpha|K_S\rangle,
\qquad
c_-
=
\langle K_S|X_\alpha|K_L\rangle.
\end{equation}

Thus the coefficients \(c_\pm\) arise from weak-interaction-induced
off-diagonal matrix elements of the projected stress tensor. In their
absence the effective operator is diagonal,
\begin{equation}
X_\alpha
=
c_0\mathbf 1
+
c_3\sigma_3,
\end{equation}
and no \(\omega\)-effect is generated.
We now project ${\cal T}_\alpha$ onto the two-dimensional neutral-kaon
subspace:
\begin{equation}
X_\alpha
=
P {\cal T}_\alpha P,
\end{equation}
where
\begin{equation}
P
=
|K_S\rangle\langle K_S|
+
|K_L\rangle\langle K_L|.
\end{equation}

The resulting operator acts only on
\(\{|K_S\rangle,|K_L\rangle\}\)
and therefore admits the Pauli decomposition
\begin{equation}
X_\alpha
=
c_0\mathbf 1
+
c_1\sigma_1
+
c_2\sigma_2
+
c_3\sigma_3 .
\end{equation}

Equivalently,
\begin{equation}
X_\alpha
=
c_0\mathbf 1
+
c_3\sigma_3
+
c_+\sigma_+
+
c_-\sigma_- ,
\end{equation}
with
\begin{equation}
c_+
=
c_1-i c_2,
\qquad
c_-
=
c_1+i c_2.
\end{equation}

The coefficients \(c_\pm\) are precisely the off-diagonal matrix elements
\begin{equation}
c_+
=
\langle K_L|X_\alpha|K_S\rangle,
\qquad
c_-
=
\langle K_S|X_\alpha|K_L\rangle .
\end{equation}

Only these transition matrix elements contribute to the
\(\omega\)-effect.


\subsection{Origin and form of off-diagonal terms}

As explained in Section~\ref{sec:interaction}, off-diagonal components of the
meson--graviton coupling arise because the physical eigenstates $K_S$ and $K_L$
are weak-interaction mixtures of the flavour states $K^0$ and $\bar K^0$. When
the stress tensor is projected onto the propagation-eigenstate basis, any
operator that is off-diagonal in flavour space (such as the strangeness-changing
part of the effective weak Hamiltonian) acquires non-zero transition matrix
elements between $K_S$ and $K_L$. The effective meson operator therefore takes
the form $X_{\rm eff} = X_{\rm diag}+x_+\sigma_++x_-\sigma_-$, where tracing
over the gravitational bath also produces a renormalisation of the diagonal part.
The off-diagonal part $x_+\sigma_++x_-\sigma_-$ is essential: generating
$|K_SK_S\rangle$ from $|K_SK_L\rangle$ requires the flip $K_L\to K_S$ (via
$\sigma_-$), and generating $|K_LK_L\rangle$ requires $K_S\to K_L$ (via
$\sigma_+$). Any coupling with $x_+=x_-=0$ gives $\mathcal{C}_{AS}^{(K)}=0$.

\subsection{Explicit calculation}

In the interaction picture with $\Delta\lambda=\lambda_L-\lambda_S
=\Delta m_K-\frac{i}{2}\Delta\Gamma_K$,
\begin{equation}
X(t) = x_0\mathbf{1} + x_3\sigma_3
+ x_+e^{+i\Delta\lambda t}\sigma_+ + x_-e^{-i\Delta\lambda t}\sigma_-.
\end{equation}
Writing $X_a = X(t_a)$ with matrix entries $(A_a,B_a;C_a,D_a)$ in the
$(K_S,K_L)$ basis, a direct calculation gives
\begin{align}
\mathcal C_{AS}^{(K)}(t_1,t_2)
&= \tfrac12\bigl[A_1B_2-B_1A_2-C_1D_2+D_1C_2\bigr]
\nonumber\\
&=\tfrac12\Bigl[
(x_0+x_3)x_+\bigl(e^{+i\Delta\lambda t_2}-e^{+i\Delta\lambda t_1}\bigr)
+(x_0-x_3)x_-\bigl(e^{-i\Delta\lambda t_1}-e^{-i\Delta\lambda t_2}\bigr)
\Bigr].
\end{align}
This confirms the selection rule: $\mathcal{C}_{AS}^{(K)}=0$ if $x_+=x_-=0$.
The full contribution to $\omega$ is
\begin{equation}
\omega(t)
= -\sum_\alpha g_\alpha^{(1)}g_\alpha^{(2)}\,e^{i\phi_\alpha}
\sinh r_\alpha\cosh r_\alpha
\int_0^t dt_1\int_0^{t_1}dt_2\;
e^{-i\Omega_\alpha(t_1+t_2)}\,\mathcal C_{AS}^{(K)}(t_1,t_2),
\end{equation}
where the squeezed bath supplies the anomalous correlator and the kaon sector
supplies the flavour-changing matrix element; $\omega$ is produced only by their
product.

\section{Derivation of the Graviton Pair-Production Kernel}
\label{app:Jkernel}

We derive the effective tensor $J^{\mu\nu\rho\sigma}$ appearing in the
squeezing kernel; for more details see~\cite{Dorlis:2025amf,Dorlis:2025zzz}.

\subsection{Quadratic graviton source from the axion action}

Expanding the axion action $S_b = -\frac12\int d^4x\sqrt{-g}[g^{\alpha\beta}
\partial_\alpha b\,\partial_\beta b+\mu_b^2 b^2]$ to second order in
$h_{\mu\nu}$, with the axion evaluated on the classical cloud
$b_{\rm cl}(t,\mathbf{x}) \simeq b_0 R_{nlm}(r)Y_{lm}(\theta,\phi)
\cos(\omega_b t-m\phi+\delta)$, yields $S_b = S_b^{(0)}+S_b^{(1)}+S_b^{(2)}+\cdots$.
The linear term describes the ordinary graviton--axion coupling. The term
quadratic in $h_{\mu\nu}$ is
\begin{equation}
S_b^{(2)}
=\frac{\kappa^2}{2}\int d^4x\sqrt{-\bar g}\;
h_{\mu\nu}\,J^{\mu\nu\rho\sigma}(x)\,h_{\rho\sigma},
\label{eq:quadratic_action_appendix}
\end{equation}
where, in the approximately flat region far from the horizon,
\begin{align}
J^{\mu\nu\rho\sigma}
=\;&
\tfrac12\eta^{\mu\rho}\partial^\nu b\,\partial^\sigma b
+\tfrac12\eta^{\mu\sigma}\partial^\nu b\,\partial^\rho b
+\tfrac12\eta^{\nu\rho}\partial^\mu b\,\partial^\sigma b
+\tfrac12\eta^{\nu\sigma}\partial^\mu b\,\partial^\rho b
\nonumber\\
&-\tfrac14\eta^{\mu\nu}(\partial^\rho b\,\partial^\sigma b+\partial^\sigma b\,\partial^\rho b)
-\tfrac14\eta^{\rho\sigma}(\partial^\mu b\,\partial^\nu b+\partial^\nu b\,\partial^\mu b)
\nonumber\\
&+\tfrac18(\eta^{\mu\nu}\eta^{\rho\sigma}
-2\eta^{\mu\rho}\eta^{\nu\sigma}-2\eta^{\mu\sigma}\eta^{\nu\rho})
[(\partial b)^2+\mu_b^2 b^2],
\label{eq:J_explicit_appendix}
\end{align}
with all axion fields on the classical configuration.

\subsection{Squeezing kernel and resonance structure}

Expanding the graviton field in transverse-traceless modes and substituting
into~\eqref{eq:quadratic_action_appendix} produces terms $a_i^\dagger a_j^\dagger$
with coefficients
\begin{equation}
\mathcal{G}_{ij}
\sim
-i\kappa^2\int d^4x\sqrt{-\bar g}\;
\Pi^{(h_i)}_{\mu\nu}(\hat{\mathbf k}_i)\,u_{k_i}(x)\,
J^{\mu\nu\rho\sigma}(x)\,
\Pi^{(h_j)}_{\rho\sigma}(\hat{\mathbf k}_j)\,u_{k_j}(x).
\label{eq:Z_kernel_appendix}
\end{equation}
Since the cloud oscillates as $b_{\rm cl}\sim e^{-i\omega_b t}+e^{+i\omega_b t}$,
the quadratic source contains pieces $\sim e^{\pm 2i\omega_b t}$. The rotating
wave approximation selects the resonant condition $\Omega_1+\Omega_2\simeq 2\omega_b$,
corresponding to coherent annihilation of two axion quanta into a correlated
graviton pair. Accordingly, $Z_{ij}$ is peaked on anti-collinear pairs
$\mathbf k_j\simeq-\mathbf k_i$. The Takagi decomposition of this kernel
determines the independent squeezed supermodes that drive the $\omega$-effect.

\section{Axion-Cloud Bound-State Profile}
\label{app:axion_cloud_profile}

In this appendix we summarise the origin of the radial profile
$R_{nlm}(r)$ appearing in the classical axion-cloud background
\begin{equation}
b_{\rm cl}(t,\mathbf{x})
\simeq
b_0\,
R_{nlm}(r)\,
Y_{lm}(\theta,\phi)\,
\cos\!\left(
\omega_b t
-
m\phi
+
\delta
\right).
\end{equation}

The axion field satisfies the massive Klein--Gordon equation on the Kerr
background,
\begin{equation}
\left(
\Box-\mu_b^2
\right)b=0.
\end{equation}
Separating variables as
\begin{equation}
b(t,r,\theta,\phi)
=
e^{-i\omega t}
e^{im\phi}
S_{lm}(\theta;a\omega)
R_{lm}(r),
\end{equation}
one obtains the spin-zero Teukolsky radial equation
\begin{equation}
\frac{d}{dr}
\left(
\Delta \frac{dR}{dr}
\right)
+
\left[
\frac{
\left(
(r^2+a^2)\omega-am
\right)^2
}{\Delta}
-
\mu_b^2 r^2
-
\Lambda_{lm}
\right]R
=
0,
\end{equation}
where
\begin{equation}
\Delta=r^2-2Mr+a^2,
\end{equation}
and $\Lambda_{lm}$ is the angular separation constant.

The quasi-bound-state boundary conditions are ingoing behaviour at the
horizon,
\begin{equation}
R
\sim
e^{-i(\omega-m\Omega_H)r_*},
\end{equation}
and exponential decay at spatial infinity,
\begin{equation}
R
\sim
e^{-kr},
\qquad
k=\sqrt{\mu_b^2-\omega^2}.
\end{equation}
Here \(r_*\) denotes the Kerr tortoise coordinate, defined by
\begin{equation}
\frac{dr_*}{dr}
=
\frac{r^2+a^2}{\Delta},
\qquad
\Delta
=
r^2-2Mr+a^2
=
(r-r_+)(r-r_-),
\label{eq:tortoise_def}
\end{equation}
where
\begin{equation}
r_\pm
=
M\pm\sqrt{M^2-a^2}
\end{equation}
are the outer and inner Kerr horizon radii.

Integrating Eq.~\eqref{eq:tortoise_def} gives
\begin{equation}
r_*
=
r
+
\frac{2Mr_+}{r_+-r_-}
\ln|r-r_+|
-
\frac{2Mr_-}{r_+-r_-}
\ln|r-r_-|.
\label{eq:tortoise_explicit}
\end{equation}

The tortoise coordinate maps the exterior region
\(r\in(r_+,\infty)\) to
\(r_*\in(-\infty,\infty)\), with
\begin{equation}
r\rightarrow r_+
\quad\Longrightarrow\quad
r_*\rightarrow -\infty,
\end{equation}
and
\begin{equation}
r\rightarrow\infty
\quad\Longrightarrow\quad
r_*\rightarrow +\infty.
\end{equation}

In terms of \(r_*\), the asymptotic behaviour of the radial mode
functions is particularly simple. Near the horizon,
\begin{equation}
u_{\omega\ell m}
\sim
e^{-i(\omega-m\Omega_H)r_*},
\qquad
\Omega_H
=
\frac{a}{r_+^2+a^2},
\end{equation}
while at spatial infinity
\begin{equation}
u_{\omega\ell m}
\sim
e^{\pm i\omega r_*}.
\end{equation}

In the non-relativistic regime
\begin{equation}
\alpha_g
=
GM\mu_b
\ll 1,
\end{equation}
the system reduces to a gravitational atom. The radial functions are then
hydrogenic:
\begin{equation}
R_{nl}(r)
=
N_{nl}
\left(
\frac{2r}{na_0}
\right)^l
e^{-r/(na_0)}
L_{n-l-1}^{2l+1}
\!\left(
\frac{2r}{na_0}
\right),
\end{equation}
with Bohr radius
\begin{equation}
a_0
=
\frac{1}{\mu_b\alpha_g}
=
\frac{1}{GM\mu_b^2}.
\end{equation}
The corresponding bound-state frequency is
\begin{equation}
\omega_b
\simeq
\mu_b
\left(
1-\frac{\alpha_g^2}{2n^2}
\right).
\end{equation}

The dominant superradiant level is usually the
\begin{equation}
n=2,
\qquad
l=1,
\qquad
m=1
\end{equation}
state. Its hydrogenic radial profile is
\begin{equation}
R_{211}(r)
=
\frac{1}{2\sqrt6\,a_0^{3/2}}
\left(
\frac{r}{a_0}
\right)
e^{-r/(2a_0)}.
\end{equation}

Thus, in the approximation used in the main text, the classical axion
cloud may be written as
\begin{equation}
b_{\rm cl}(t,\mathbf{x})
\simeq
b_0
R_{211}(r)
Y_{11}(\theta,\phi)
\cos\!\left(
\omega_b t-\phi+\delta
\right),
\end{equation}
with $R_{211}(r)$ given above. In a full Kerr calculation, this
hydrogenic profile should be replaced by the exact quasi-bound solution
of the radial Teukolsky equation.

\section{Takagi Decomposition and Graviton Supermodes}
\label{app:takagi}
 
This appendix provides a self-contained discussion of the Takagi
decomposition and its application to the multimode squeezed graviton state
used in the main text.
The graviton creation and annihilation operators satisfy the canonical
commutation relations
\begin{equation}
  [a_i, a_j^\dagger] = \delta_{ij}, \qquad
  [a_i, a_j] = [a_i^\dagger, a_j^\dagger] = 0.
\end{equation}
The multimode squeezed vacuum is defined by
\begin{equation}
  |\Psi\rangle
  =
  \mathcal{N}\exp\!\left[
    \tfrac{1}{2}\sum_{ij}
    \left(
      \mathcal G_{ij}\,a_i^\dagger a_j^\dagger
      -
      \mathcal G_{ij}^*\,a_i a_j
    \right)
  \right]|0\rangle,
  \label{eq:squeezed_vac}
\end{equation}
 where $\mathcal G_{ij}$ is the squeezing kernel. Bose symmetry requires
\begin{equation}
  \mathcal G_{ij} = \mathcal G_{ji}.
\end{equation}
The matrix $\mathcal G$ is therefore a complex symmetric matrix.
In the axion-cloud problem, $\mathcal G_{ij}$ is obtained by projecting the
quadratic graviton source onto pairs of outgoing graviton modes:
\begin{equation}
  \mathcal G_{ij}
  \sim
  -i\int d^4x\;J(x)\;u_i(x)\;u_j(x),
\end{equation}
where $u_i(x)$ denotes the graviton mode functions and $J(x)$ is the
effective source generated by the axion cloud.

 \paragraph{Takagi's theorem:}
 Every complex symmetric matrix \(\mathcal G\) admits a Takagi decomposition
\cite{Houde:2024mkj}
\begin{equation}
  \mathcal G = U R U^T ,
  \label{eq:takagi}
\end{equation}
where \(U\) is unitary and
\begin{equation}
  R=\mathrm{diag}(r_1,r_2,\ldots),
  \qquad
  r_\alpha\ge 0 .
\end{equation}
The quantities \(r_\alpha\) are the Takagi singular values. They are real
and non-negative, and therefore should be interpreted as the squeezing
amplitudes.

In component form,
\begin{equation}
  \mathcal G_{ij}
  =
  \sum_\alpha
  U_{i\alpha}\,
  r_\alpha\,
  U_{j\alpha}.
  \label{eq:Gij_decomp}
\end{equation}
Although the diagonal matrix \(R\) contains no phases, the kernel
\(\mathcal G_{ij}\) is in general complex because the unitary matrix \(U\) is
complex. Thus the phases of the squeezed supermodes are encoded in the
Takagi vectors, not in the singular values \(r_\alpha\).

Equivalently, one may rephase the Takagi mode operators by
\begin{equation}
  \mathbb b_\alpha
  \longrightarrow
  e^{i\theta_\alpha}\mathbb b_\alpha .
\end{equation}
Under this redefinition the phase convention of the corresponding Takagi
vector changes, while the physical kernel \(\mathcal G\) is unchanged. Therefore
only phase-invariant combinations, such as the anomalous correlators
entering the reduced meson dynamics, have invariant physical meaning.

With the Takagi supermodes defined by
\begin{equation}
  \mathbb b_\alpha
  =
  \sum_i U^*_{i\alpha} a_i ,
\end{equation}
the squeezing operator factorises into independent single-supermode
squeezers,
\begin{equation}
|\Psi\rangle
=
\prod_\alpha
\exp\left[
\frac12 r_\alpha
\left(
\mathbb b_\alpha^{\dagger 2}
-
\mathbb b_\alpha^2
\right)
\right]
|0\rangle ,
\end{equation}
in a convention where the squeezing phases have been absorbed into the
definition of the \(\mathbb b_\alpha\). If one chooses instead to work
with rephased operators in which the squeezing phases are explicit, one
may write
\begin{equation}
|\Psi\rangle
=
\prod_\alpha
\exp\left[
\frac12 r_\alpha
\left(
e^{i\phi_\alpha}\mathbb b_\alpha^{\dagger 2}
-
e^{-i\phi_\alpha}\mathbb b_\alpha^2
\right)
\right]
|0\rangle .
\end{equation}
The two forms are equivalent; they differ only by a phase convention for
the Takagi modes.
Since $U$ is unitary, these satisfy
\begin{equation}
  [\mathbb{b}_\alpha, \mathbb{b}_\beta^\dagger] = \delta_{\alpha\beta},
\end{equation}
and therefore the $\mathbb{b}_\alpha$ define a new canonical bosonic basis.
These are the \emph{Takagi supermodes}. Physically, they are the linear
combinations of graviton modes that are independently squeezed by the
source.

Substituting~\eqref{eq:Gij_decomp} into the exponent
of~\eqref{eq:squeezed_vac} gives
\begin{align}
  \sum_{ij} \mathcal G_{ij}\,a_i^\dagger a_j^\dagger
  &= \sum_\alpha r_\alpha e^{i\phi_\alpha}
     \left(\sum_i U_{i\alpha}\,a_i^\dagger\right)
     \left(\sum_j U_{j\alpha}\,a_j^\dagger\right)
  \nonumber\\
  &= \sum_\alpha r_\alpha e^{i\phi_\alpha}\,\mathbb{b}_\alpha^{\dagger 2}.
\end{align}
Similarly,
\begin{equation}
  \sum_{ij} \mathcal G_{ij}^*\,a_i a_j
  =
  \sum_\alpha r_\alpha e^{-i\phi_\alpha}\,\mathbb{b}_\alpha^2.
\end{equation}
The squeezed vacuum therefore factorises as
\begin{equation}
  |\Psi\rangle
  =
  \prod_\alpha
  S_\alpha(r_\alpha,\phi_\alpha)\,|0_\alpha\rangle,
\end{equation}
where
\begin{equation}
  S_\alpha(r_\alpha,\phi_\alpha)
  =
  \exp\!\left[
    \tfrac{1}{2}
    \left(
      r_\alpha e^{i\phi_\alpha}\,\mathbb{b}_\alpha^{\dagger 2}
      -
      r_\alpha e^{-i\phi_\alpha}\,\mathbb{b}_\alpha^2
    \right)
  \right]
\end{equation}
is the single-mode squeezing operator. Thus the complicated multimode
squeezed state has been reduced to a product of independent squeezed
supermodes (known as single mode squeezers).

The Takagi basis is particularly useful because all bath correlators
become diagonal. The normal-ordered correlator is
\begin{equation}
  \langle \mathbb{b}_\alpha^\dagger \mathbb{b}_\beta\rangle
  =
  \delta_{\alpha\beta}\,\sinh^2\! r_\alpha.
\end{equation}
The anomalous correlator is
\begin{equation}
  \boxed{
  \langle \mathbb{b}_\alpha \mathbb{b}_\beta\rangle
  =
  -\delta_{\alpha\beta}\,e^{i\phi_\alpha}
  \sinh r_\alpha\cosh r_\alpha.
  }
  \label{eq:anomalous}
\end{equation}
This quantity is the characteristic signature of squeezing: it vanishes
for a coherent state or thermal state but is non-zero for a squeezed
state. In the main text it is precisely this correlator that generates
the $\omega$-effect.
The bath two-point function entering the master equation is
\begin{equation}
  \langle \mathbb{b}_\alpha(t)\,\mathbb{b}_\beta(s)\rangle
  =
  -\delta_{\alpha\beta}\,
  e^{i\phi_\alpha}\sinh r_\alpha\cosh r_\alpha\,
  e^{-i\Omega_\alpha(t+s)}.
\end{equation}

\bibliographystyle{apsrev4-2}
\bibliography{Compliant}
\end{document}